\documentclass[aps,pra,twocolumn,a4paper,showpacs,superscriptaddress]{revtex4}

\usepackage{amsmath, amssymb,mathrsfs}
\usepackage{graphicx}
\usepackage{ulem}
\usepackage{nicefrac}
\usepackage[squaren]{SIunits}
\usepackage{xcolor}
\usepackage{times}
\usepackage{hyperref}
\usepackage{textcomp}
\usepackage[T1]{fontenc}

\begin{document}

\title{Quantum Measurement of Broadband Nonclassical Light Fields}

\author{P. Gr\"unwald}
\email{peter.gruenwald@uni-rostock.de}
\affiliation{AG Theoretische Quantenoptik Institut f\"ur Physik, Universit\"at Rostock, Universit\"atsplatz 3, D-18051 Rostock, Germany}

\author{D.~Vasylyev}
%\affiliation{School of Physics and Astronomy, University of St Andrews, North Haugh, St Andrews KY16 9SS, UK}
\affiliation{Bogolyubov Institute for Theoretical Physics, NAS of Ukraine, Metrolohichna 14-b, UA-03680 Kiev, Ukraine.}

\author{J. \surname{H\"aggblad}}
\affiliation{Department of Numerical Analysis, KTH Royal Institute of Technology, SE-100 44 Stockholm, Sweden}

\author{W. Vogel}
\affiliation{AG Theoretische Quantenoptik Institut f\"ur Physik, Universit\"at Rostock, Universit\"atsplatz 3, D-18051 Rostock, Germany}

\date{\today}

\newcommand{\tno} {\,^\circ_\circ\,}

\begin{abstract}
Based on the measurement of quantum correlation functions, the quantum statistical properties of spectral measurements are studied for broadband radiation fields. The spectral filtering of light before its detection is compared with the direct detection followed by the spectral analysis of the recorded photocurrents. As an example, the squeezing spectra of the atomic resonance fluorescence are studied for both types of filtering procedures. The conditions for which the detection of the nonclassical signatures of the radiation is possible are analyzed. For the considered example, photocurrent filtering appears to be the superior option to detect nonclassicality, due to the vacuum-noise effects in the optical filtering.
\end{abstract}

\maketitle
\section{Introduction}
%In standard optical experiments the filtering of the measured optical signals is usually desired. 
Filtering of optical signals plays an important role in experimental quantum optics. The  optical field under investigation always includes unwanted components, or noise, contributing to the signal due to the imperfections and losses in constituents of the optical setup or due to the environment surrounding this setup. From the quantum optical point of view, it is impossible to have no loss at all~\cite{Haus:00}. The task of the experimenter is to minimize the inaccuracies, caused by the presence of such noise, by proper filtering.

Optical filtering is a process in which certain spectral parts of the signal are suppressed due to convolution with a filter function, which represents the selecting device. The most common filters are glasses, specifically designed to transmit some definite wavelengths, which are placed in the input ports of the detectors~\cite{Edmund}.
Alternatively, electric current filters can be used in the detectors output channels~\cite{Zverev:76}. These filters are realized mainly as simple electronic pass-band filters, which can be easier controlled than optical filters. From the viewpoint of classical optics the application of both filtering techniques is equivalent. Thus electronic filtering techniques have been applied in many modern experimental setups~\cite{filterhistory}. This does not only include optical experiments, but virtually every signal analysis in which a frequency dependent input is transformed into an electric current signal,  e.g. in geophysics~\cite{Geofilter}, acoustics~\cite{Acousticfilter}, and electronic devices themselves~\cite{Digitalfilter}. 

In the quantum domain the equivalence of optical and electronic filtering is by no means obvious. On one hand the disparity arises since the relation between the optical and the photoelectric current spectra strongly depends on the statistical properties of the optical signal field to be measured~\cite{Cummins}. On the other hand,
the filtering convolutions occur at very different stages of the light-analyzing process. Optical filters act on the quantum light itself, before the detector records the radiation field. As a consequence, such a filtering process unavoidably introduces additional quantum noise effects into the signal before it is measured. The current filtering, is a purely classical procedure, which is implemented after the completion of the detection process of the radiation. Hence the current filtering does not add quantum noise to the data. However, when broadband fields are measured, in general the detectors may only integrate over parts of the radiation spectrum, so that information may be lost before the currents are spectrally analyzed. Thus, it is a cumbersome problem to identify the optimal strategy for spectral measurements of quantum light fields of a broad spectral bandwidths. It is noteworthy that the optimal type of filtering may also depend on the physical situation under study. For example, the optical 
spectral filtering may be the preferential choice for the extraction of entangled photon pairs, which could be generated in the biexciton-radiative cascade process~\cite{Akopian} or by V-type three-level systems in microcavities~\cite{Ajiki}. However, electronic photocurrent filters have been useful for the measurements of the signal to noise ratio of light with a Gaussian  statistics~\cite{Cummins} and for the quadrature-fluctuation spectroscopy with squeezed light~\cite{Kuprianov}. In the following we shall focus on spectral correlation measurements,
for which both types of filtering may be applied.

The theory of passive optical filters and their influence on correlation properties of filtered quantum light was developed in Ref.~\cite{Vogel:86,Cresser,PhysRevA.36.3803, PhysRevA.42.503, Leonardt}. This topic has become of interest more recently since methods were developed and set up to measure arbitrary field correlation functions~\cite{shchukin:200403,Blatt}. The theoretical concepts, however, have proven difficult to analyze for higher order moments. Therefore, alternative descriptions have also been studied~\cite{Laussy,Silberhorn}. Based on the above argumentation the current-filtering procedure includes the implicit filtering by the detector, which acts in a similar manner as a spectral filter, as well as the classical filtering of the current signals after detection. The latter is a purely classical process.

The aim of the present paper is to compare the spectral measurements of broadband radiation, based on optical and electronic filtering. We provide a consistent theoretical approach to treat the quantum noise effects in both techniques. Furthermore, detecting normally and time ordered field correlation functions via balanced correlation homodyning with filters preserves the ordering from the original fields. Thus, the filtered fields can be used to detect nonclassicality in the same way as for the original fields. To illustrate the results, we analyze the elementary example of the squeezing spectrum of the atomic resonance fluorescence. Our finding is that the spectral filtering light limits the ability to detect the squeezing to a greater extent than the current filtering, making the latter preferential for this setup.

The paper is organized as follows. In Sec.~\ref{spectral-chapter} we will describe the techniques for measuring the spectral correlation functions of an optically filtered radiation field. 
In Sec.~\ref{photocurrentSec} the procedure of current filtering will be analyzed. Both kinds of filtering techniques are compared in Sec.~\ref{fluorescence-chapter} for the example of the squeezing spectrum in the resonance fluorescence of a two-level atom. A summary and some conclusions are given in Sec.~\ref{summary}.

\section{Correlation Properties of spectrally Filtered Light}\label{spectral-chapter}

From a mathematical point of view, the intrinsic spectral properties of a light field under study are recovered by a Fourier analysis of the signals obtained in the time domain.
In classical optics this procedure is straightforward. In quantum optics, however, the application of the spectral analysis is a more sophisticated problem, because of the time- and normal-ordering prescriptions of the field operators in the measured  correlation functions together with the related quantum noise effects~\cite{PhysRevA.42.503}.

In order to recover the information about the spectral properties of measured light, one may send the light beam through a frequency sensitive device, before detection. In classical physics, the spectrally filtered field is expressed by a convolution integral of the signal field with a filter response function. The quantum theory of photodetection of optically filtered light contains additional difficulties, due to the quantum noise effects introduced by the filtering procedure~\cite{Vogel:86,Cresser,PhysRevA.36.3803, PhysRevA.42.503, Leonardt}.  Therefore, a careful analysis of correlation properties must be performed for the filtered optical radiation fields. 

\begin{figure}[h]
\includegraphics[width=7cm]{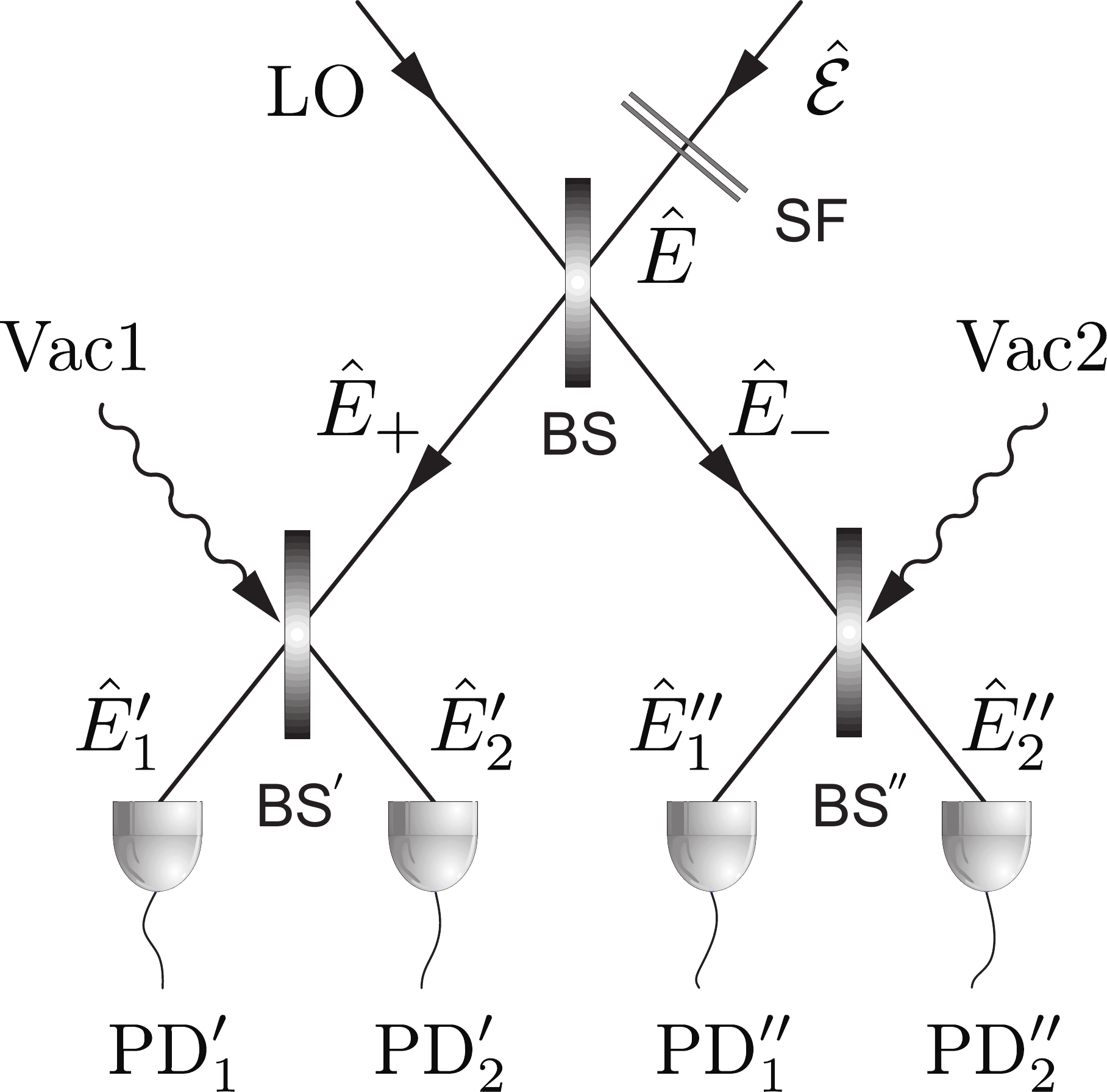}
\caption{\label{fourport_filter} The setup for four-detector correlation measurements. The signal field $\mathcal{\hat E}$ is filtered by passing through the spectral filter ($\rm SF$) and then it is mixed with the local oscillator ($\rm LO$) by a beamsplitter ($\rm BS$). The resulting field components  $\hat E_\pm$ pass through two beamsplitters $\rm BS'$ and $\rm BS''$, and are detected by four photodetectors ($\rm PD'_1, \dots $).}
\end{figure}

In Ref.~\cite{shchukin:200403} a universal measurement scheme has been proposed to measure the quantum correlation functions of light. We will briefly recall the results and refer to the paper for details. A simple example of such a setup is shown in Fig.~\ref{fourport_filter} if one neglects the spectral filter (SF). The scheme can be extended by adding more beamsplitters and detectors. It records normally-ordered intensity correlation functions $\Gamma_\ell^{(k)}$ of the light field  $\hat{\mathcal E}$, superimposed with the local oscillator (LO). The specific form of these correlation functions in our scenario will be discussed later on. These correlations are then combined in a binomial sum 
\begin{equation}\label{binom_sum}
F^{(k)} = \sum^{k}_{\ell = 0} (-1)^{k-\ell} \binom{k}{\ell} \Gamma_\ell^{(k)}\propto\langle:\hat{\mathcal X}_\varphi^k:\rangle,
\end{equation}
which is proportional to the $k$-th moment of the field quadrature $\hat{\mathcal X}_\varphi$. Herein, $2k$ is the total number of detectors and $\ell$ is the number of detectors chosen on the left side of the first beamsplitter (BS).
In this section we extend this scheme by a spectral filter (SF), thereby changing the signal field from $\hat{\mathcal E}$ to $\hat E$, in order to describe the  measurement of filtered broadband light fields. 

%%%%%%%%%%%%%%%%%%%%%%%%%%%%%%%%%%%%%%%%%%%%%%%%%%%%%%%%%%%%%%%%%%%%%%%%%%%%%%%
\subsection{Spectral filtering of light with a single filter}
Let us consider the measurement scheme proposed in~\cite{shchukin:200403} and restrict it to the case of four photodetectors, see Fig.~\ref{fourport_filter}. In this case one measures the second-order intensity correlation functions of the signal field superimposed with the LO, cf. Eq.~(\ref{binom_sum}) for $k{=}2$. This is sufficient for the detection of the squeezing spectrum. The filter in this scheme must be carefully chosen; when we add more spectral filters we need to make sure that we preserve the possibility to combine the measured data in a binomial form as in Eq.~(\ref{binom_sum}).

The original signal field will be labeled $\mathcal{\hat E}$. After transmission through the SF the resulting field $\hat E$ is a convolution  of the unfiltered field with the filter function $ T_\text{f}$ plus some (vacuum) noise field $\hat E_{\rm n}$.
Afterwards, the filtered field is superimposed with the LO via the BS and reads as \cite{PhysRevA.42.503}: 
\begin{align}\label{LOBS}
	\hat E_{\pm}^{(+)} (t)=& \frac{e^{i \phi_{\pm}}}{\sqrt{2}}\bigg[ \int d t'T_\text{f}(t{-}t')  \hat{\mathcal{E}}^{(+)}(t') \nonumber\\
	&+ \hat E^{(+)}_{\rm n} (t) {\pm} i \mathcal{\hat
	E}^{(+)}_{\text{LO}}(t) \bigg],\\
	\hat E_{\pm}^{(-)} (t)=&\left[\hat E_{\pm}^{\left(+\right)} (t)\right]^{\dagger},
\end{align}
where the upper indices $+(-)$ refer to positive(negative) frequency components of the fields, whereas the lower indices $+(-)$ refer to transmitted(reflected) parts of the incident light by the first beamsplitter  (cf. Fig.~\ref{fourport_filter}).
The two phases $\phi_\pm$ that correspond to the  fields $\hat E_\pm$ satisfy the constraint $\phi_+-\phi_-=\pi/2$.

Finally, after propagation through the other two beamsplitters $\rm BS'$ and $\rm BS''$, the fields at the photodetectors are
\begin{gather}
\hat E_{j}^{(\pm)'} = \frac{e^{\pm i \phi_j}}{\sqrt{2}}\left( \hat
  E^{(\pm)}_+ + \mathcal{\hat E_{\text{vac1}}} \right),\qquad i=1,2\\
\hat E_{j}^{(\pm)''} = \frac{e^{\pm i \phi_j}}{\sqrt{2}}\left( \hat
  E^{(\pm)}_- + \mathcal{\hat E_{\text{vac2}}} \right), \qquad i=1,2,
\end{gather}
where $\phi_{1,2}$ are the phase differences caused by the beamsplitters. The terms $\mathcal{\hat E_{\text{vac1,2}}}$ describe the vacuum contributions in the unused input ports, which are eliminated by the  normal- and time-ordering of the field correlation functions~\cite{PhysRevA.36.3803}.
Here it has been assumed that all the beamsplitters are symmetric, 50:50 ones. As usual in homodyne measurements, the LO is a strong coherent field with amplitude $E_{\rm LO}$, such that the operator nature of the $\rm LO$-field  plays no role in the observed correlation functions. Hence, the result is the same if we use a classical approximation for the $\text{LO}$,
\begin{equation}\label{coherent_state_LO}
\mathcal{\hat E}^{(-)}_\text{LO}(t) {=} E_\text{LO}
e^{i(\omega_\text{LO} t {-} \phi_{\text{LO}})}, \qquad  \mathcal{\hat
  E}^{(+)}_\text{LO} {=}\left[\mathcal{\hat E}^{(-)}_\text{LO} \right]^\ast .
\end{equation}
Consequently, only the signal field shows quantum effects in the measured quantities.

Let us define the following analogs of the photon number operator (cf.~\cite{shchukin:200403}):
\begin{equation}
\begin{split}
&\mathcal{\hat N}_{\pm} = \hat E^{(-)}_\pm\hat E^{(+)}_\pm\\
&\quad=\frac{1}{2}\Bigg[ \int d  t'_1 d  t'_2
T_\text{f}^*(t{-}t'_1) T_\text{f}(t{-}t'_2)
\mathcal{\hat E}^{(-)}(t'_1) \mathcal{\hat E}^{(+)}(t'_2)\\
&\qquad\qquad {+}\hat E^{(-)}_{\rm n}\hat E^{(+)}_{\rm n}{+}\hat E^{(-)}\hat E^{(+)}_{\rm n}{+}\hat E^{(-)}_{\rm n}\hat E^{(+)}\\
&\qquad\qquad\qquad\qquad\qquad{+} E^2_\text{LO}\pm E_\text{LO}\bigl(
{\hat X_\varphi}{+}{\hat X_{\rm n,\varphi}}\bigr)\Bigg],\label{NN}
\end{split}
\end{equation}
 where $\varphi=\varphi_\text{LO} + \pi/2$ and
\begin{align}\label{EqChi}
{\hat X}_\varphi =& \hat{\tilde E}^{(+)} e^{-i \varphi} + \hat{\tilde E}^{(-)} e^{i \varphi},\\
{\hat X}_{\rm n,\varphi} =& \hat{\tilde E}^{(+)}_{\rm n} e^{-i \varphi} + \hat{\tilde E}^{(-)}_{\rm n} e^{i \varphi},\nonumber\\
\hat{\tilde E}^{(\pm)} =& \hat{E}^{(\pm)}e^{\pm i \omega_\text{LO}t},\quad\hat{\tilde E}^{(\pm)}_{\rm n} = \hat{E}^{(\pm)}_{\rm n}e^{\pm i \omega_\text{LO}t},\label{spectral_slowly}\\
\hat{E}^{\left(+\right)} =& \int  d  t' T_\text{f}(t-t')  \hat{\mathcal{E}}^{(+)}(t')+\hat E_\text n^{(+)}.
\end{align}
Here and in the following we indicate the slowly varying field amplitudes via a tilde.
Using the definition~(\ref{NN}), we calculate the field correlation functions similar to those in~\cite{shchukin:200403}, cf. Eq.~(\ref{binom_sum}) with $k=2$. 
For $\ell$ ($0 {\leq} \ell {\leq} 2$) photodetectors on the left
side of the setup in Fig.~\ref{fourport_filter} and $2-\ell$ on the right side, we get the correlation functions
\begin{equation}\label{photodetector_correlation}
\Gamma_{\ell}^{(2)}=2^{-2} \left\langle \tno \mathcal{ \hat N}^\ell_+ \mathcal{\hat
 N}^{2-\ell}_- \tno \right\rangle \qquad 0 \leq \ell \leq 2.
\end{equation}
Combining  Eqs.~(\ref{binom_sum}), (\ref{NN}) and  (\ref{photodetector_correlation})  
we obtain for the spectral filtered version of the quantity $F^{(2)}$ defined in Eq.~(\ref{binom_sum}) the expression
\begin{align}\label{spectral_f_function}
&F^{(2)}_\text{spectral} = 2^{-2} \sum^{2}_{\ell=0} (-1)^{2-\ell} \binom{2}{\ell} \left\langle \tno \mathcal{ \hat N}^\ell_+ \mathcal{\hat N}^{2-\ell}_- \tno \right\rangle\nonumber\\
&\quad{=}\frac{1}{2^{2}} \Big\langle \tno \left( \mathcal{\hat N}_+ {-}\mathcal{\hat N}_- \right)^2 \tno \Big\rangle= \frac{E^2_\text{LO}}{2^2} \left\langle \tno \hat{X}_\varphi^2 \tno \right\rangle.
\end{align}
Here $\tno\ldots\tno $ denotes the normal and time ordering prescription~\cite{Vogel:06}. The ordering allows the application of the binomial summation, which leads to higher order moments of $\hat{X}_\varphi$.
Using Eq.~(\ref{EqChi}), we may write Eq.~(\ref{spectral_f_function}) explicitly as
 \begin{multline}\label{spectral_expanded}
F^{(2)}_\text{spectral} {=} \frac{E^2_\text{LO}}{2^2}\int  d  t'_1 \int  d  t'_2 \\
{\times}\bigg\langle\tno
  \prod_{i=1}^2\bigg[T_\text{f}(t{-}t'_i)
\mathcal{\hat E}^{(+)}(t'_i)e^{i(\omega_\text{LO} t - \varphi)}\\
+
   T_\text{f}^*(t{-}t'_i)\mathcal{\hat E}^{(-)}(t'_i)e^{-i(\omega_\text{LO} t - \varphi)}\bigg]\tno \bigg\rangle .
\end{multline}
This formula generalizes the result of Ref.~\cite{shchukin:200403} for the case of spectrally  filtered  radiation fields.

Performing the Fourier transformation  of Eq.~(\ref{spectral_f_function}) with respect to the phase $\varphi$, we are able to reconstruct the moments of field operators according to
\begin{equation}
\int_0^{2\pi}\hspace{-0.3cm} d \varphi F^{(n+m)}_\text{spectral} e^{-i(n-m)\varphi} \propto \left\langle
  \tno \hat{\tilde E}^{(-)n}\hat{\tilde E}^{(+)m} \tno \right\rangle, \label{eq.recon_mom}
\end{equation}
with $m$ and $n$ being integers.
For the case $k=2$ Eq.~(\ref{eq.recon_mom})  yields
\begin{eqnarray}
\int_0^{2\pi}\hspace{-0.3cm} d \varphi F^{(2)}_\text{spectral} e^{-i2\varphi} &=& \frac{\pi}{2}
E^2_{\text{LO}} \left\langle \tno
  \hat{\tilde E}^{(-)2} \tno \right\rangle,\\
\int_0^{2\pi}\hspace{-0.3cm} d \varphi F^{(2)}_\text{spectral} &=& \pi
E^2_{\text{LO}} \left\langle \tno
\hat{\tilde E}^{(-)}\hat{\tilde E}^{(+)} \tno \right\rangle,\\
\int_0^{2\pi}\hspace{-0.3cm} d \varphi F^{(2)}_\text{spectral} e^{i2\varphi} &=& \frac{\pi}{2}
E^2_{\text{LO}} \left\langle \tno \hat{\tilde E}^{(+)2} \tno \right\rangle. \label{eq.moments}
\end{eqnarray}
These moments, when expressed in terms of the signal fields, are for the case of Eq.~(\ref{eq.moments}) of the form
\begin{eqnarray}\label{one_spectral_filter_moments}
\left\langle\tno \hat{\tilde E}^{(+)2}\tno\right\rangle&=& 
\int d  t_1 \int d  t_2 T_\text{f}(t{-}t_1)T_\text{f}(t{-}t_2)\nonumber\\
&&\times e^{2i\omega_\text{LO}t}\left\langle\tno\hat{\mathcal E}^{(+)}(t_1)\hat{\mathcal E}^{(+)}(t_2) \tno\right\rangle.
\end{eqnarray}
Hence, we obtained the connection between the incident light fields, the filter functions and the fields at the detector. 
%%%%%%%%%%%%%%%%%%%%%%%%%%%%%%%%%%%%%%%%%%%%%%%%%%%%%%%%%%%%%

%%%%%%%%%%%%%%%%%%%%%%%%%%%%%%%%%%%%%%%%%%%%%%%%%%%%%%%%%%%%%
\subsection{Spectral filtering of light with two filters}
Let us turn to the case of two optical filters applied within the measurement setup. Calculating the correlations of optical fields with different frequencies allows us to resolve the squeezing spectrum. 
Again, the filters must be configured in a manner to allow the binomial summation.

\begin{figure}[h]
\includegraphics[width=7.5cm]{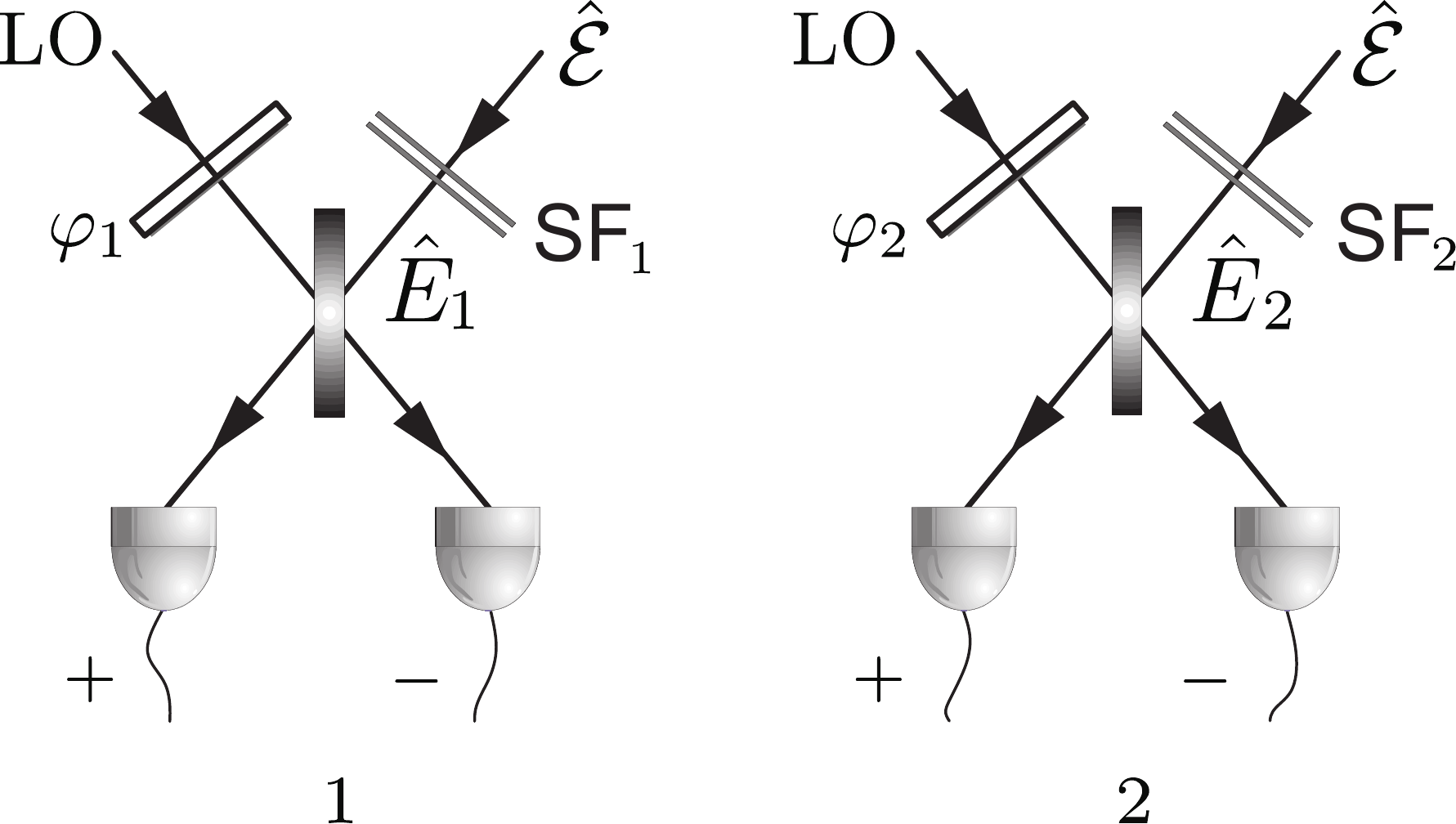}
\caption{\label{twin_doubleport_filter}The four-detector measurement scheme for correlations of electromagnetic waves of different frequencies and phases. The signal field $\mathcal{\hat E}$ in the $j$-th arm of the setup ($j=1,2$) is passing through the spectral filter ${\rm SF}_j$. Then it is mixed with the phase-controlled ${\rm LO}$. The resulting beams are detected by the photodetectors ${\rm PD}_\pm$. The outcomes of the photodetectors are correlated.}
\end{figure}

The setup is given in Fig.~\ref{twin_doubleport_filter}. The signal field $\mathcal{\hat E}$  is split in two equal parts and each one passes one of two different homodyning setups. At the spectral filters $\rm SF_1$ and $\rm SF_2$ the signal field $\hat{\mathcal E}$ transforms into the fields $\hat E_{1}$ and $\hat E_{2}$. These fields are then mixed with two $\rm LO$s with different phases $\varphi_1$ and $\varphi_2$ and then impinge on the four detectors. The detected fields are
\begin{equation}
\hat E^{(+)}_{j,\pm} = \frac{e^{i \phi_\pm}}{\sqrt 2}\left( \hat E^{(+)}_j \pm
i \hat{ \mathcal{E}}^{(+)}_{j,\text{LO}}\right),
\end{equation}
where each detector is numbered by the  index $\{j,\pm\}, j=1,2$, which refers to the corresponding subdevice in Fig.~\ref{twin_doubleport_filter}. The filtered fields $\hat E_j$ are related to the unfiltered ones as
\begin{equation}
\hat E^{(+)}_j= \int  d  t'_j T_{\text{f}_j}(t-t'_j) \mathcal{\hat E}^{(+)}(t_j') + E^{(+)}_{j, \rm n} ,\label{eq.18}
\end{equation}
where the response functions $T_{\text{f}_j}(t-t'_j)$ describe the action of the filter devices. For the local oscillator field in a coherent state, the photon number operators read  as
\begin{equation}\label{twin_setup_photon_number_operator}
\begin{split}
\mathcal{\hat N}_{j,\pm}=& \hat E^{(-)}_{j,\pm} \hat E^{(+)}_{j,\pm} {=} \frac{1}{2}\left(\hat
E^{(-)}_j \hat E^{(+)}_j {+}\hat E^{(-)}_{j,{\rm n}} \hat E^{(+)}_{j,{\rm n}} \right.\\
&\hspace{-0.2cm}\left.{+}\hat E^{(-)}_j \hat E^{(+)}_{j,{\rm n}}{+}\hat E^{(-)}_{j,{\rm n}} \hat E^{(+)}_j{+} E^2_{j,\text{LO}} {\pm} E_{j,\text{LO}}{\hat X}_{j,\varphi} \right),
\end{split}
\end{equation}
with
\begin{equation}
\hat{X}_{j,\varphi} = \bigl(\hat{\tilde E}_j^{(+)}+\hat{\tilde E}^{(+)}_{j,{\rm n}} \bigr) e^{-i \varphi_j} +\text{H.c.} %\hat{\tilde E}_i^{(-)} e^{i \varphi_i}
\end{equation}
 and tilde denotes the slowly-varying field component, e.g. $\hat{\tilde E}^{(\pm)}_j {=} \hat E^{(\pm)}_j e^{\pm i\omega_{j,\text{LO}}t}$. Note also that $\varphi_j{=}\varphi_{j,\text{LO}}{+}\pi/2$. 

Now we need to correlate the detected signals from both filter arms. Consequently, we may chose two indices $\ell$ and $m$ with $0\leq \ell,m\leq k$ for the left and the right setup, respectively. The normally ordered correlation functions of the photodetectors are
\begin{align}\label{G11}
\Gamma^{(1,1)}_{\ell,m} 
= \left\langle\tno \hat{\mathcal N}_{1,+}^{\ell}\hat{\mathcal N}_{1,-}^{1-\ell}\hat{\mathcal N}_{2,+}^m\hat{\mathcal N}_{2,-}^{1-m}\tno \right\rangle.
\end{align}
The  upper double indices  of $\Gamma^{(d_1,d_2)}_{\ell,m}$ indicate the depth levels  of the homodyning measurement in each arm of the setup  and are equal to half of the numbers $k_j$ ($j{=}1,2$) of detectors placed  in each arm  of the  setup. Since in our case both indices are equal to one we can  use this setup to measure second order correlation functions of two frequencies.

Using Eq.~(\ref{G11}), analogously to Eq.~(\ref{binom_sum}), we define
\begin{equation}
F^{(1,1)}_\text{spectral} = \sum^1_{\ell=0} \sum_{m=0}^1 (-1)^{1-\ell} (-1)^{1-m}\; \Gamma^{(1,1)}_{\ell,m}.
\end{equation}
Applying the binomial formula we obtain
\begin{equation}\label{eq.fspectral11}
 \begin{split}
	F^{(1,1)}_\text{spectral}=& \left\langle \tno \left( \hat{\mathcal N}_{1,+} {-} \hat{\mathcal
N}_{1,-} \right) \left( \hat{\mathcal N}_{2,+} {-} \hat{\mathcal N}_{2,-}
\right) \tno \right\rangle\\
	=& E_\text{LO}^2\left\langle \tno \hat{X}_{1,\varphi_1}\hat{X}_{2,\varphi_2} \tno \right\rangle=E_{\rm LO}^2\int\!\! d  t_1'\int\!\! d  t_2'\\
	&\hspace{-0.1cm}\times\hspace{-0.1cm}\left\langle\tno\prod_{j=1}^2\Bigg[T_{\text f_j}(t-t_j')\hat{\mathcal E}^{(+)}(t_j')e^{i(\omega_{j,{\rm LO}}t-\varphi_j)}\right.\\
	&\hspace{-0.1cm}+\left.T_{\text f_j}^*(t-t_j')\hat{\mathcal E}^{(-)}(t_j')e^{-i(\omega_{j,{\rm LO}}t-\varphi_j)}\Bigg]\tno\right\rangle,
\end{split}
\end{equation}
which can be used for the reconstruction of the field operator moments, similarly as in Eq.~(\ref{eq.recon_mom}).
After performing two dimensional Fourier-transformation, we arrive at
\begin{equation}\label{two_spectral_moments}
\begin{split}
&\left\langle\tno \hat{\tilde E}_1^{(\pm)} \hat{\tilde E}^{(\pm)}_2
  \!\tno\!\right\rangle\\& \quad=\int d  t_1 \!\int d  t_2
  T_{\text{f}_1}^{(\pm)}(t{-}t_1)T_{\text{f}_2}^{(\pm)}(t{-}t_2)\\
&\times e^{i(\pm\, \omega_{1,\text{LO}} \, \pm \,
  \omega_{2,\text{LO}})t }\left\langle\tno 
  \hat{\mathcal{E}}^{(\pm)}(t_1)\hat{\mathcal{E}}^{(\pm)}(t_2)
  \tno\right\rangle,
\end{split}
\end{equation}
where $T_{\text{f}}^{+}(t){=}T_{\text{f}}(t)$ and $T_{\text{f}}^{-}{=}[T_{\text{f}}^{+}]^\ast$. This formula can be compared with Eq.~(\ref{one_spectral_filter_moments}) for the corresponding expression for one filter frequency.

\section{Photocurrent filtering}\label{photocurrentSec}

The other major technique of spectral detection used in experiments is based on current filtering. In this method the photoelectric current generated from the light field incident on the detector is filtered.
The obvious advantages are that the light field itself is not modified by the filter and the technical process of current filtering is much easier  controlled than optical selective devices. Furthermore, as we have mentioned above, current filtering is a classical process, since the current is already the output of the detection.

%%%%%%%%%%%%%%%%%%%%%%%%%%%%%%%%%%%%%%%%%%%%%%%%%%%%%%%%%%%%%
\subsection{Photocurrent filtering with one filter frequency}

\begin{figure}[h]
\includegraphics[width=7.5cm]{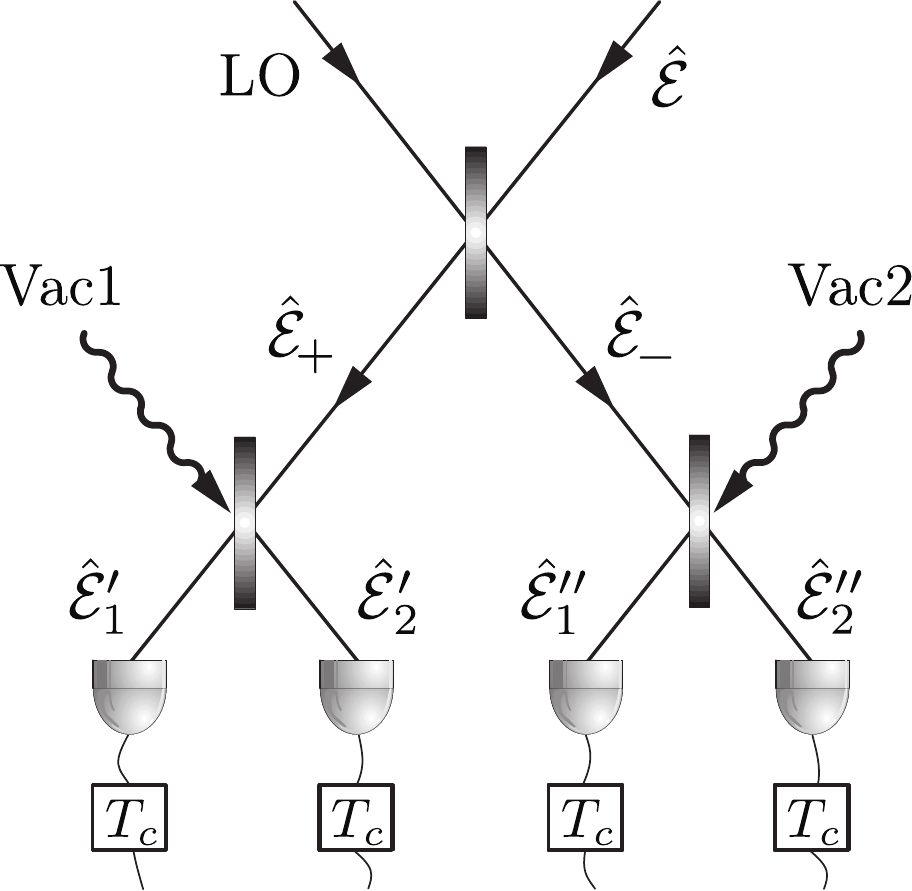}
\caption{\label{fourport}Four-detector setup with current filtering. The outcomes of the photodetection measurement are filtered by the current filters $T_c$.}
\end{figure}

Let us consider the four-detector setup, shown in Fig.~\ref{fourport}. Instead of the optical spectral filters, four electronic filters act on the photocurrents. In the following, we analyze the measurement scheme in more detail.

Following the procedure given in Ref.~\cite{Vogel:06}, that describes the detector operation based on quantum and classical statistics, we introduce the $\hat\Gamma$-functions 
\begin{equation}
\hat\Gamma(t,\Delta t) = N \!\int^{t{+}\Delta t}_{t} \hspace{-0.7cm} d \tau \int^{t{+}\Delta
  t}_{t} \hspace{-0.7cm}  d \tau' S(\tau {-} \tau') \mathcal{\hat E}^{(-)}(\tau)\mathcal{\hat
  E}^{(+)}(\tau').
\end{equation}
It corresponds to the observable measured by a single detector. These functions hold for $N$ identical atoms in a point-like detector setup irradiated by light within the time interval $t,t+\Delta t$; $S(\tau)$ is the detector response function. In the situation where the bandwidth of the field is much  narrower than the detector bandwidth, the detector response is usually approximated by a delta function, the so called `broad-band-detector approximation'. Here this is not justified and we keep $S(\tau)$ in the form of a general function.

Now we implement the results of~\cite{Vogel:06}, and calculate the correlation of two detectors, indicated by 1 and 2, measuring over the same interval $\Delta t$, but from different initial times $t_1$ and $t_2$,
\begin{align}
\overline{n(t_1,\Delta t)n(t_2,\Delta t)} &=\hspace{-0.3cm} \sum_{m_{1,2}=0}^{\infty}\hspace{-0.3cm}m_1m_2P_{m_1,m_2}(t_1,\Delta t,t_2,\Delta t)\nonumber\\ 
&=\! \left\langle \tno\hat\Gamma^{(1)}(t_1,\Delta t) \hat\Gamma^{(2)}(t_2,\Delta t) \tno\right\rangle \! ,\label{eq.26}
\end{align}
where $n (t_j,\Delta t)$ is the number of `clicks' in the detector  $j$. Here, $P_{m_1,m_2}(t_1,\Delta t,t_2,\Delta t)$ is the joint probability of emission of $m_1$ photoelectrons within the time interval $t_1,t_1+\Delta t$ in detector 1 and $m_2$ photoelectrons within $t_2,t_2+\Delta t$ in detector $2$. 
Eq.~(\ref{eq.26}) is equivalent to the corresponding expression for one detector in a case of non-overlapping time intervals $t_1,t_1+\Delta t$ and $t_2,t_2+\Delta t$. For two detectors such an overlap is not relevant. The only correlation stems from the fact, that the same light field is incident on both detectors, given by the $\hat \Gamma$-operators.

The  photocurrent generated in an electron multiplying detector can be described as $i(t)=ge\, n(t,\Delta t)/\Delta t$ with $g$ being the gain factor, which we assume to be constant.  We thus obtain two-time current-correlation function of the form
\begin{equation}\label{current_correlation}
\overline{i_1(t_1)i_2(t_2)} = \frac{g^2 e^2}{(\Delta t)^2} \left\langle\! \tno\!
  \hat\Gamma^{(1)}(t_1,\Delta t) \hat\Gamma^{(2)}(t_2,\Delta t) \!\tno\! \right\rangle.
\end{equation}
The correlation function for the  filtered currents, 
\begin{equation}
i_{\text f}(t){=}\int  d  t' T_\text{c}(t{-}t') i(t'),
\end{equation}
is calculated to be
\begin{align}
\overline{i_{1\text f}(t_1)i_{2\text f}(t_2)} =& \frac{g^2 e^2}{(\Delta t)^2} \int
 d  t'_1 T_\text{c}(t_1{-}t'_1)\! \int  d  t'_2 T_\text{c}(t_2{-}t'_2)\nonumber\\ 
&\times \left\langle\! \tno\! \hat\Gamma^{(1)}(t'_1,\Delta t) \hat\Gamma^{(2)}(t'_2,\Delta t)\! \tno \!\right\rangle.\label{eq.32}
\end{align}

Now we can turn to the special scheme in Fig.~\ref{fourport} and define the appropriate $\hat \Gamma$-operators as
\begin{align}
\hat\Gamma'_j(t,\Delta t) &{=} N \!\int^{t{+}\Delta t}_{t} \hspace{-0.7cm} d \tau \int^{t{+}\Delta
  t}_{t} \hspace{-0.7cm}  d \tau' S(\tau {-} \tau') \mathcal{\hat E}_j^{(-)'}(\tau)\mathcal{\hat
  E}_j^{(+)'}(\tau'),\\
\hat\Gamma''_j(t,\Delta t) &{=} N\! \int^{t+\Delta t}_{t}\hspace{-0.7cm}  d \tau
  \int^{t{+}\Delta t}_{t} \hspace{-0.7cm}  d \tau' S(\tau {-} \tau') \mathcal{\hat E}_j^{(-)''}(\tau)\mathcal{\hat E}_j^{(+)''}(\tau').
\end{align}
One prime denotes here the left arm of the detector setup, whereas double prime denotes the right arm (cf. Fig.~\ref{fourport}).

The detected fields $\hat{\mathcal E}^{'}_j$ and $\hat{\mathcal E}^{''}_j$  are expressed through  linear combinations of fields $\hat{\mathcal E}_-$ and $\hat{\mathcal E}_+$ and vacuum contributions. 
Defining 
\begin{align}
&\hat\Gamma_\pm(t'_j,\Delta t) {=} \frac{N}{2}\hspace{-0.3cm} \int\limits^{t'_j{+}\Delta t}_{t'_j}\hspace{-0.3cm} d \tau \hspace{-0.1cm}\int\limits^{t'_j{+}\Delta
  t}_{t'_j}\hspace{-0.3cm}  d \tau' S(\tau {-} \tau') \mathcal{\hat E}_\pm^{(-)}(\tau)\mathcal{\hat
  E}_\pm^{(+)}(\tau'),\label{eq.Gamma_pm}
\end{align}
as a correlation function for the field, that would be detected right after the signal and $\rm LO$  fields pass the first beamsplitter, one can show, after some straightforward algebra, that  $\langle\tno\hat\Gamma_+^\ell\hat\Gamma^{2-\ell}_-\tno\rangle$=
$\langle\tno\hat\Gamma_{i}^{'^{\ell}}\hat\Gamma^{''^{2-\ell}}_{j}\tno\rangle$, with $i,j=1,2$ and $\ell=0,1,2$.
Then it is easy to see, that the (equal time) current correlation functions for our system can be written as
\begin{equation*}
\begin{split}
&\overline{i_{+}(t)^\ell \; i_{-}(t)^{2{-}\ell}}\\
&{=} \frac{g^2 e^2}{(\Delta t)^2}\! \int
 d  t'_1 T_\text{c}(t{-}t'_1)\! \int  d  t'_2 T_\text{c}(t{-}t'_2)
\left\langle\! \tno\! \hat\Gamma_+^{\ell} \hat\Gamma_-^{2{-}\ell}\! \tno\! \right\rangle,
\end{split}
\end{equation*}
and the subscript $\pm$ refers to the corresponding fields/detectors on the left ($+$) and right ($-$) side of the first beamsplitter.

Having obtained the expression for the correlation functions we are interested in, we construct the $F^{(k)}_\text{current}$ function [cf. Eq.~(\ref{binom_sum})]
\begin{align}
F^{(k)}_\text{current}&=\sum^k_{\ell = 0}(-1)^{k-\ell} \binom{k}{\ell} \overline{n_{+}^\ell n_{-}^{k-\ell}},\quad n_\pm=\frac{\Delta t}{ge}i_\pm.\\
\end{align}
For our setup with $k=2$ this expression reduces to 
% \begin{widetext}
% \begin{eqnarray}
%   F^{(2)}_\text{current} &=& \sum^2_{\ell = 0}(-1)^{2-\ell} \binom{2}{\ell} \overline{n_{+}^\ell
% n_{-}^{2-\ell}} = \overline{n_-^2} - \overline{2n_+ n_-} + \overline{n_+^2}\nonumber\\
% % 	&=& \int d  t_1 T_\text{c}(t{-}t_1) \int  d  t_2T_\text{c}(t{-}t_2)\left\langle \tno\left(\hat\Gamma_+ {-} \hat\Gamma_-\right)^2 \tno\right\rangle\nn\\
% 	&=& \int
%  d  t_1 T_\text{c}(t{-}t_1) \int  d  t_2 T_\text{c}(t{-}t_2) \left\langle
%    \tno\prod_{i=1}^2 \left( \hat\Gamma_+(t_i) {-} \hat\Gamma_-(t_i) \right)\tno\right\rangle. \label{eq.F2curr}
% \end{eqnarray}
% \end{widetext}

\begin{align}
  F^{(2)}_\text{current} =& \sum^2_{\ell = 0}(-1)^{2-\ell} \binom{2}{\ell} \overline{n_{+}^\ell n_{-}^{2-\ell}} \nonumber\\
	=& \int d t_1 T_\text{c}(t{-}t_1) \int  d  t_2 T_\text{c}(t{-}t_2) \nonumber\\
	&\times\left\langle\tno\prod_{i=1}^2 \left( \hat\Gamma_+(t_i) {-} \hat\Gamma_-(t_i) \right)\tno\right\rangle. \label{eq.F2curr}
\end{align}
In turn, the fields $\hat{\mathcal E}_{\pm}$ in Eq.~(\ref{eq.Gamma_pm}) for a symmetric beamsplitter are linear combinations of signal and $\rm LO$ fields, leading to
\begin{equation*}
\hat{\mathcal E}^{(-)}_\pm \hat{\mathcal E}^{(+)}_\pm {=}
\frac{1}{2}\bigg[\hat{\mathcal E}^{(-)} \hat{\mathcal E}^{(+)} {+} \hat{\mathcal
  E}^{(-)}_\text{LO} \hat{\mathcal E}^{(+)}_\text{LO} \pm i\hat{\mathcal E}^{(-)}
 \hat{\mathcal E}^{(+)}_\text{LO} {\mp} i\hat{\mathcal E}^{(-)}_\text{LO} \hat{\mathcal
  E}^{(+)} \bigg].
\end{equation*}
With the help of this relation, the difference of two correlation functions in Eq.~(\ref{eq.F2curr}) can be reduced to
\begin{multline}
\hat\Gamma_+(t_i) {-} \hat\Gamma_-(t_i) {=}
\frac{NE_{\text{LO}}}{2}\int^{t_i+\Delta t}_{t_i} \!\!\!\!\!  d \tau
\int^{t_i+\Delta t}_{t_i} \!\!\!\!\!  d \tau'
S(\tau{-}\tau') \\ 
\times \left[ \mathcal{\hat E}^{(-)}(\tau)e^{-i\omega_\text{LO}\tau'
    {+}i\varphi} {+} \mathcal{\hat{ E}}^{(+)}(\tau')e^{i\omega_\text{LO}\tau {-}i \varphi}\right],\label{eq.1CurFilDiff}
\end{multline}
where Eq.~\eqref{coherent_state_LO} has been used and $\varphi = \varphi_{\text{LO}} + \pi/2$. Thus, the full expression for the $F^{(2)}_\text{current}$ function becomes
\begin{equation}\label{filtered_current:full_expression_F}
\begin{split}
& F^{(2)}_\text{current} =\frac{N^2E^2_{\text{LO}}}{2^2} \int
 d  t_1 T_\text{c}(t{-}t_1) \int  d  t_2 T_\text{c}(t{-}t_2)\\
&\quad\times \bigg\langle\!\!\tno\!\! \prod^{2}_{j=1} \int^{t_j{+}\Delta t}_{t_j} \!\!\!\!\!  d \tau_j
\int^{t_j{+}\Delta t}_{t_j} \!\!\!\!\!  d \tau'_j S(\tau_j{-}\tau'_j) \\
&\quad\times\left[ \mathcal{\hat{E}}^{(-)}(\tau_j)e^{-i\omega_\text{LO}\tau'_j {+} i \varphi} {+}
  \mathcal{\hat{E}}^{(+)}(\tau'_j)e^{i\omega_\text{LO} \tau_j {-}i \varphi}\right]\!\! \tno \!\!\bigg\rangle.
\end{split}
\end{equation}

The obtained result can be compared with Eq.~(\ref{spectral_expanded}) for the radiation filtering case. One should note, that the essential difference between radiation and current filtering now becomes obvious. Namely, the spectral filtering process is performed before the quantum mechanical averaging procedure, whereas the current filtering acts on the averaged light field. At the same time, one should note that the detector response function acts similar to an optical spectral filter now. Hence, for both methods, a certain degree of optical filtering is unavoidable when dealing with broadband fields.

%%%%%%%%%%%%%%%%%%%%%%%%%%%%%%%%%%%%%%%%%%%%%%%%%%%%%%%%%%%%%
\subsection{Filtered current using two filter frequencies}
\begin{figure}[h]
\includegraphics[width=7.5cm]{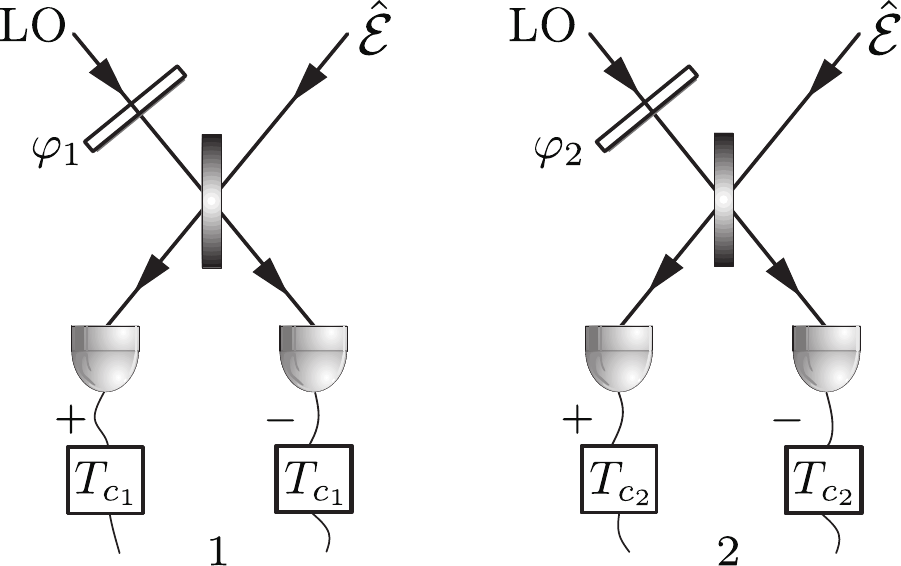}
\caption{\label{twin_doubleport}Modified scheme of Fig.~\ref{twin_doubleport_filter} without radiation filtering but with the current filtering devices $T_{c_j}$ for the $j$-th arm of the setup ($j=1,2$).}
\end{figure}

Extending the concept of current filtering to the case of two current filters tuned on different frequencies, we adopt the scheme in Fig.~\ref{twin_doubleport}. In order to  construct the $F_\text{current}$-function we note the following useful relation for the field moments being detected 
\begin{equation}\label{eq.f11current}
\begin{split}
\sum_{\ell{=}0}^1 \sum_{m=0}^1 (-1)^{1{-}\ell}(-1)^{1{-}m} \overline{
  n_{1,+}^\ell \; n_{1,-}^{1-\ell}  n_{2,+}^m \; n_{2,-}^{1-m} } \\
=
\overline{n_{1,-} \; n_{2,-}} {-} \overline{n_{1,+} \;
    n_{2,-}} {-} \overline{n_{1,-} \; n_{2,+}} {+} \overline{n_{1,
+} \; n_{2,+}}.
\end{split}
\end{equation}
Here the indices $1,2$ refer to photons detected in different arms of the setup. The $F_\text{current}$-function  which involves filtered currents can be expressed with the help of Eq.~(\ref{eq.f11current}) in terms of the $\hat\Gamma$-operators as
\begin{eqnarray}
F^{(1,1)}_\text{current} &=& \int d  t_1 T_{\text{c}_1}(t{-}t_1) \!\int d  t_2
T_{\text{c}_2}(t{-}t_2) \nonumber\\
&&{\times} \left\langle\!\tno \!\hat\Gamma_{1,-}\hat\Gamma_{2,-} {-}
  \hat\Gamma_{1,+}\hat\Gamma_{2,-}\right.\nonumber\\
&&-\left. \hat\Gamma_{1,-}\hat\Gamma_{2,+} {+} \hat\Gamma_{1,+}\hat\Gamma_{2,+}\tno \right\rangle.\label{eq.F11Cur}
\end{eqnarray}

The sum inside the normal- and time-ordering in Eq.~(\ref{eq.F11Cur}) can be evaluated analogously to Eq.~(\ref{eq.1CurFilDiff}), yielding
\begin{multline}
\prod_{j=1}^2\left[ \hat\Gamma_{j,+}(t_j) {-} \hat\Gamma_{j,-}(t_j)\right]
    \\
{=} NE_\text{LO}\prod_{j=1}^2\int^{t_j+\Delta t}_{t_j} \!\!\!\!\! d \tau
\int^{t_j+\Delta t}_{t_j} \!\!\!\!\! d \tau' S(\tau{-}\tau')\\
\times
\Big[ \hat{\mathcal
  E}^{(-)}(\tau) e^{-i(\omega_{j,\text{LO}} \tau' - \varphi_j)} {+}
\hat{\mathcal E}^{(+)}(\tau')e^{i(\omega_{j,\text{LO}} \tau - \varphi_j)}
\Big],
\end{multline}
where $\varphi_j = \varphi_{j,\text{LO}} + \pi/2$. The full expression
for $F^{(1,1)}_\text{current}$ follows as
\begin{equation}
\begin{split}
&F^{(1,1)}_\text{current} = N^2 E^2_\text{LO} \int  d  t_1 T_{\text{c}_1}(t{-}t_1) \int  d  t_2 T_{\text{c}_2}(t{-}t_2) \\
&\times \Bigg\langle\!\!\tno\!\! \prod_{j=1}^2\int^{t_j{+}\Delta t}_{t_j} \hspace{-0.5cm} d \tau_j\int^{t_j{+}\Delta t}_{t_j} \hspace{-0.5cm} d \tau'_jS(\tau_j {-}\tau'_j)\\
&\times\left[ \hat{\mathcal
  E}^{(-)}(\tau_j) e^{-i(\omega_{j,\text{LO}} \tau'_j {-} \varphi_j)} {+}
\hat{\mathcal E}^{(+)}(\tau'_j)e^{i(\omega_{j,\text{LO}} \tau_j {-} \varphi_j)} \right]\!\!
\tno\!\!\Bigg\rangle. \label{two_filters_for_current}
\end{split}
\end{equation}
Again Eq.~(\ref{two_filters_for_current}) can be compared with the corresponding expression~(\ref{eq.fspectral11}) for the  radiation filtering case.

One should note, that both methods of spectral filtering can be applied in one experimental setup as well. For the description of this case, one will have to combine the formalisms for the two cases above. This calculation is straightforward but lengthy.
Otherwise it is interesting to compare the two methods and raise the question under which conditions the different filterings are useful from the viewpoint of an experiment.
%%%%%%%%%%%%%%%%%%%%%%%%%%%%%%%%%%%%%%%%%%%%%%%%%%%%%%%%%%%%%

\section{Application to Resonance Fluorescence}\label{fluorescence-chapter}

As an example for our calculations, let us now consider the resonance fluorescence from a driven two-level atom as a source field. We are interested in nonclassical properties  of the resonance fluorescence, namely the squeezing phenomenon  predicted in Refs~\cite{WallsZoller}, \cite{Mandel} and then  verified experimentally \cite{LuBali}. Based on the influence of the two filtering processes, we discuss, which method is preferable for this specific quantum optical problem.

The light field of interest is emitted by a free two-level atom (with the transition frequency $\omega_{21}$), which in turn is irradiated with a resonant laser field of the same frequency $\omega_{\rm L}= \omega_{21}$. The total emission field can be written as
\begin{align}
\hat{\vec{\mathcal E}}(\vec r,t)=&\hat{\vec{\mathcal E}}^{(+)}(\vec r,t)+\hat{\vec{\mathcal E}}^{(-)}(\vec r,t),\\
\hat{\vec{\mathcal E}}^{(+)}(\vec r,t)=&\hat{\vec{\mathcal E}}^{(+)}_{\text{free}}(\vec r,t) + \hat{\vec{\mathcal E}}^{(+)}_{\text{s}}(\vec r,t),\label{EEsEf}\\
\hat{\vec{\mathcal E}}^{(-)}(\vec r,t)=&\left(\hat{\vec{\mathcal E}}^{(+)}(\vec r,t)\right)^\dagger,
\end{align}
where the source field is given by
\begin{equation}
\hat{\vec{\mathcal E}}^{(+)}_{\text{s}}(\vec r,t) = \vec g(\vec r - \vec r_{\text A}) \hat
A_{12}\left(t- | \vec r - \vec r_{\text A} |/c\right).\label{eq.sourcefield}
\end{equation}
Herein, $\hat A_{ab} = |a\rangle \langle b|$ are the atomic flip operators ($\{a,b\}=1,2$ refer to  ground and excited state of an atom, respectively) evaluated at the retarded times $t_R{=}t{-} | \vec r {-} \vec r_{\text A} |/c$, with $\vec r_{\text A}$ being the position of the atom and $\vec g$ relates atomic operators to the field quantities.
We assume, that the free field  is in the vacuum state at the detectors. Hence, only the source field part of Eq.~(\ref{eq.sourcefield}) is observed in measurements of time- and normally-ordered correlation functions. For simplicity, in the following we shall denote it by $\hat{\vec{\mathcal E}}$.

\subsection{The Bloch equations}

In order to evaluate the correlation functions for our filtered correlations as in Eqs.~(\ref{eq.fspectral11}),(\ref{two_filters_for_current}) we need explicit information about the incident field $\hat{\vec{\mathcal E}}$. For the basic methods to deal with atomic resonance fluorescence we refer to~\cite{Vogel:06}. We start with the optical Bloch equations that describe the time evolution of the radiating atom,
\begin{align}
\dot\sigma_{22} &= -\Gamma_1\sigma_{22} - \frac{1}{2}i\Omega_\text{R} \tilde\sigma_{21} + \frac{1}{2}i\Omega_\text{R} \tilde\sigma_{12},\label{eq.Bloch22}\\
\dot\sigma_{11} &= \Gamma_1\sigma_{22} + \frac{1}{2}i\Omega_\text{R} \tilde\sigma_{21} - \frac{1}{2}i\Omega_\text{R} \tilde\sigma_{12},\\
\dot{\tilde\sigma}_{21} &= -\Gamma_2\tilde\sigma_{21} + \frac{1}{2}i\Omega_\text{R} \left(\sigma_{11} - \sigma_{22} \right),\\
\dot{\tilde\sigma}_{12} &= -\Gamma_2\tilde\sigma_{12} - \frac{1}{2}i\Omega_\text{R} \left(\sigma_{11} - \sigma_{22} \right)\label{eq.Bloch12},
\end{align}
where $\sigma_{ab}=\langle\hat A_{ba}\rangle$ are the density matrix elements with slowly varying  diagonal elements. The off-diagonal elements are split into a fast oscillating term $\propto \exp(\pm i\omega_\text Lt)$ and a slowly varying term $\tilde\sigma_{ab}$, $a\neq b$. Moreover, $\Omega_\text{R}$ is the Rabi frequency and $\Gamma_a$, $a=1,2$ are the energy and phase damping rates, respectively. 

Using the quantum regression theorem~\cite{Vogel:06}, \cite{PR129.2342}, we define
\begin{equation}\label{G-ab}
G_{ab}(\tau)=\left\langle \hat A_{ba}(\tau) \hat A_{12}(0) \right\rangle, \qquad \tau\geq0.
\end{equation}
The correlation functions $G_{ab}$ obey the same Bloch equations as $\sigma_{ab}$, but 
the initial conditions are given by
\begin{equation}
G_{ab}(0)=\delta_{a1}\sigma_{2b}.
\end{equation}
As we deal with a continuous-wave scenario, the explicit values of the initial conditions for $G_{ab}$ follow from the steady state values of $\sigma_{ab}$.
The system of differential equations~(\ref{eq.Bloch22})-(\ref{eq.Bloch12}) can be solved more easily by reformulating the correlation functions as Laplace integrals.
We define
\begin{equation}\label{S-ab}
\tilde S_{ab}(s)=\int_0^{+\infty}\!\!\!\!\! d \tau e^{-s\tau}\tilde G_{ab}(\tau)
\end{equation}
as the Laplace-transform of the slowly varying $\tilde G_{ab}$ functions (cf.~\cite{Vogel:06}), which leads to algebraic equations in place of Eqs.~(\ref{eq.Bloch22})-(\ref{eq.Bloch12}). 

The relevant solutions for the $\tilde S_{ab}$-functions are
\begin{equation}\label{S-12}
	\tilde S_{12}(s) = \frac{\sigma_{22}(\infty)}{s+\Gamma_2} - \tilde S_{21}(s)
\end{equation}
and
\begin{equation}\label{S-21}
\begin{split}
\tilde S_{21}(s)&=\frac{i\Omega_\text{R}}{2s(s+\Gamma_2)}\Bigr[\tilde\sigma_{21}(\infty)\Bigl.\\
&\qquad\Bigl.{-}\frac{is\Omega_\text{R}\sigma_{22}(\infty)+\Omega_\text{R}^2\tilde\sigma_{21}(\infty)}{\left[(s+\Gamma_1)(s+\Gamma_2)+\Omega_\text{R}^2 \right]}\Bigr],
\end{split}
\end{equation}
which are expressed by the steady-state solutions of the density matrix elements,
\begin{align}
&\sigma_{22}(\infty)=\frac{1}{2}\frac{\Omega_\text{R}^2}{\Gamma_1\Gamma_2+\Omega_\text{R}^2},\\
	&\tilde\sigma_{21}(\infty)=\frac{i}{2}\frac{\Gamma_1\Omega_\text{R}}{\Gamma_1\Gamma_2+\Omega_\text{R}^2}.
\end{align}
The solutions of the system of Bloch-equations are further used for the  calculation of the electromagnetic field correlation functions. Here we intend
to calculate the normally ordered squeezing spectrum as defined in~\cite{Collet:84},
\begin{equation}\label{eq.Sq-spec}
	S_{\text{sq}}(\omega)=\frac{1}{2\pi}\int d \tau e^{i\omega\tau}\left\langle \tno \Delta\hat{\overrightarrow{\mathcal E}}(\tau)\Delta\hat{\overrightarrow{\mathcal E}}(0) \tno \right\rangle,
\end{equation}
where we use $\Delta \hat{\overrightarrow{ \mathcal E}}=\hat{\overrightarrow{ \mathcal E}}-\langle\hat{\overrightarrow{ \mathcal E}}\rangle$. The squeezing spectrum (\ref{eq.Sq-spec}) follows from Eqs.~(\ref{S-12}) and (\ref{S-21}) by inserting Eqs.~(\ref{eq.sourcefield}), (\ref{G-ab}), and (\ref{S-ab}). We shall now discuss the squeezing spectrum for both  spectral and current filtering of resonance fluorescence light.
%%%%%%%%%%%%%%%%%%%%%%%%%%%%%%%%%%%%%%

%%%%%%%%%%%%%%%%%%%%%%%%%%%%%%%%%%%%%%%
\subsection{The squeezing spectrum of filtered light}
As special filters used in the detection scheme, we choose Lorentz-type filter functions with different filter frequencies $\omega_{\rm f_i}$ ($i{=}1,2$), but equal pass bandwidths $\Gamma_{\rm f}$. For details on the filters we refer to the Appendices~\ref{AppA}, \ref{AppB}. Using the measurement scheme of Fig.~\ref{twin_doubleport_filter} with Lorentzian filters $\rm SF_1$ and $\rm SF_2$, we reconstruct the spectral function $F_{\rm spectral}^{(1,1)}$ by means of Eq.~(\ref{eq.fspectral11}), which can be related to the field moments (\ref{two_spectral_moments}) by  two-dimensional Fourier transform.   By  Eq.~(\ref{eq.Sq-spec}), we can express the squeezing spectrum as a function of $\Delta\omega{=}\omega_{\rm f_2}{-}\omega_{\rm f_1}$.
We characterize squeezing in the form 
\begin{align}
	S_{\text{sq}}^{\text{max}}(\Delta\omega)=&\frac{2}{\pi\Gamma_\text{f}|\mathbf g|^2}\int \hspace{-0.1cm} d\tau\, \text{Re} \Big\{ \left\langle \tno \hat{E}^{(-)}_1(\tau)\hat{E}^{(+)}_2(0) \tno \right\rangle \nonumber\\
	&\hspace{1.7cm}- \left\langle \tno \hat{E}^{(+)}_1(\tau)\hat{E}^{(+)}_2(0) \tno \right\rangle \Big\}e^{i\Delta\omega\tau}\\
	=&\frac{2}{\pi}\text{Re}\left[ \frac{\sigma_{22}(\infty)}{\Gamma_2 + \Gamma_\text{f} -i\Delta\omega} - \tilde S_{21}(\Gamma_\text{f}-i\Delta\omega)\right],\label{eq.maxsq}
\end{align}
which  is considered for those phases of the field, for which squeezing is maximally pronounced.

\subsubsection{Idealized filtering of light}
\begin{figure}[h]
\includegraphics[width=7.5cm]{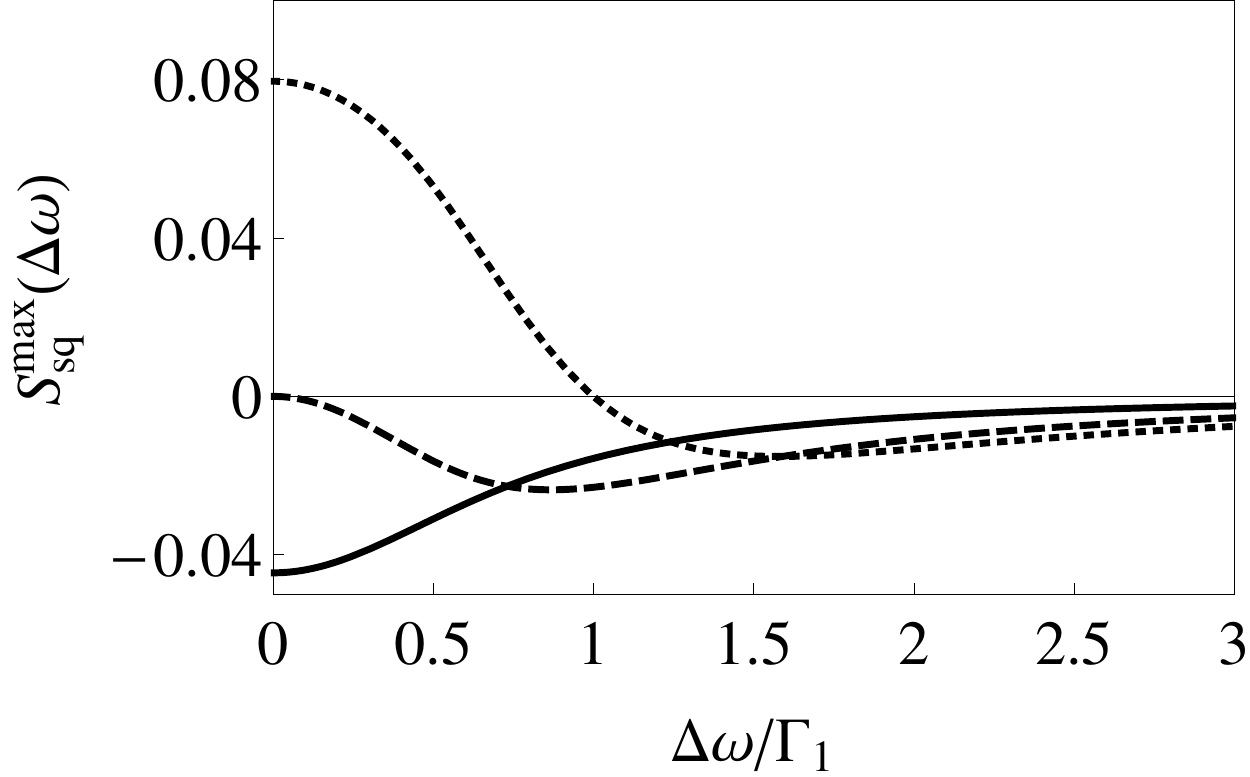}
\caption{\label{correlation_collet_1} The spectrum $S_{\text{sq}}^{\text{max}}$ for the maximally squeezed phase, for different values of the Rabi frequency $\Omega_\text{R}^2/\Gamma_1^2$: 1/2 (dotted), 1/4 (dashed), 1/12 (solid).}
\end{figure}

Unless mentioned otherwise, we will in the following consider the atom in the purely radiative damping regime, that is $\Gamma_1=2\Gamma_2$. For the special case when the pass bandwidth of the spectral filter goes to zero ($\Gamma_f{\rightarrow}0$), the detected squeezing spectrum in Eq.~(\ref{eq.maxsq}) coincides with the one calculated in Ref.~\cite{Collet:84}.
This spectrum is shown in Fig.~\ref{correlation_collet_1} for various values of the Rabi frequency. Squeezing is present when $S_{\text{sq}}^{\text{max}}<0$. In the $\Delta\omega$ region where this condition holds true, the fluctuations of the field are below the vacuum noise level.

For small values of $\Omega_{\rm R}^2/\Gamma_1^2$ (cf. with solid line in Fig.~\ref{correlation_collet_1}) the term $\langle\tno \hat E^{(+)}_1\hat E^{(+)}_2\tno\rangle$ contributes stronger than $\langle\tno \hat E^{(-)}_1\hat E^{(+)}_2\tno\rangle$ to the squeezing spectrum, resulting in a Lorentzian dip below the vacuum level. The maximal squeezing is obtained for $\Omega_{\rm R}^2/\Gamma_1^2=\frac{1}{12}$ in agreement with the results of Refs.~\cite{WallsZoller}, \cite{Swain}. For increasing  excitations (dotted line in Fig.~\ref{correlation_collet_1}) the spectrum of inelastically scattered light shows a pronounced peak of half-width $\Gamma_1$ centered on the driving frequency $\omega_{\rm L}$. This peak is superimposed with the Lorentzian dip of half-width $\Omega_{
\rm R}'{=}\sqrt{\Omega_{\rm R}^2{+}\frac{1}{2}\Gamma_1^2}$. In the strong-driving limit ($\Omega_{\rm R}^2{\gg}\frac{1}{2}\Gamma_1^2$) the main contribution to $S_{\text{sq}}^{\text{max}}$ stems from $\langle\tno \hat E^{(-)}_1\hat E^{(+)}_2\tno\rangle$. The spectrum $S^{\rm max}_{\rm sq}(\Delta\omega)$ shows two peaks situated at frequencies $\Delta\omega{=}{\pm}\Omega_{\rm R}'$  that correspond to the sideband peaks of the Mollow triplet~\cite{Mollow}. While squeezing for small filter detuning $\Delta\omega$ is absent in this case,  we always find some squeezing at higher detuning if $\Gamma_1>\Gamma_2$ holds. For $\Gamma_1=\Gamma_2$ the radiationless dephasing becomes as large as the energy relaxation rate, destroying all squeezing~\cite{PRA2013}. For $\Gamma_1>\Gamma_2$, we obtain a negative squeezing spectrum for
\begin{equation}
   (\Delta\omega)^2>\frac{2\Omega_\text R^2\Gamma_1}{\Gamma_1-\Gamma_2}-\Gamma_1^2.\label{eq.negSpec}
\end{equation}

\subsubsection{Realistic filtering of light}
\begin{figure}[h]
\includegraphics[width=7.5cm]{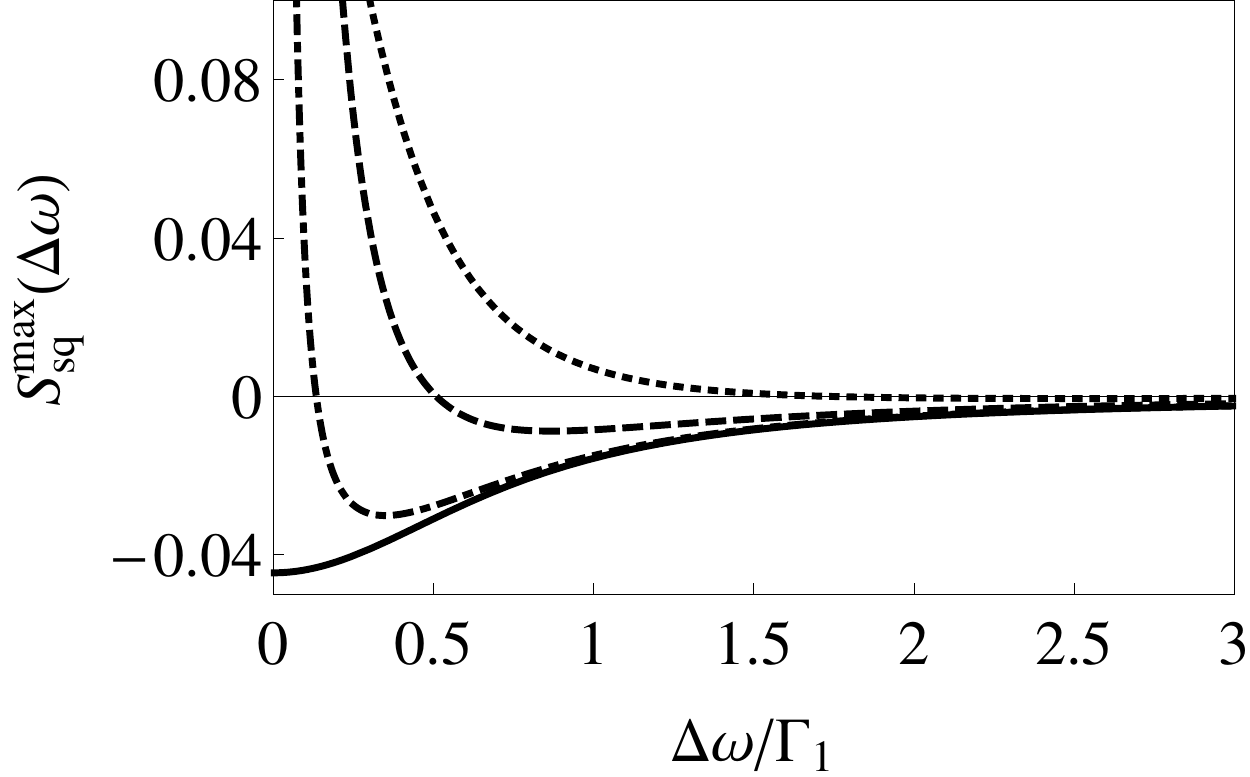}
\caption{\label{correlation_filter_1} The spectrum $S_{\text{sq}}^{\text{max}}$ for $\Omega^2_{\text R}/\Gamma_1^2=1/12$ and different values of the pass bandwidth $\Gamma_\text{f}/\Gamma$: 1/3 (dotted), 1/10 (dashed), 1/100 (dash-dotted), 0 (solid).}
\end{figure}

We now turn to the more realistic case of non-zero filter width. Fig.~\ref{correlation_filter_1} depicts the squeezing spectra for different values of $\Gamma_{\text f}/\Gamma_1$ in the case of weak pumping $\Omega_\text R^2/\Gamma_1^2=1/12$. One can clearly see, that, in contrast to the idealized filtering ($\Gamma_{\text f}=0$), realistic values of the spectral filter bandwidths significantly reduce the 
accessible squeezing effect, especially for small filter detunings $\Delta\omega$. For $\Gamma_{\text f}/\Gamma_1=1/100$, the squeezing effect is preserved for $\Delta\omega>\Gamma_2$, whereas for higher values of the filterwidth only a small squeezing effect can be observed. Increasing the filter bandwidths further quickly destroys the squeezing effect, which almost disappears already for $\Gamma_\text{f}/\Gamma_1=1/3$. However, similar to the case of idealized filtering, we find some squeezing for sufficiently large filter detuning. The former condition for negativity generalizes to
\begin{equation}
   \Gamma_1>\Gamma_2+\Gamma_\text f.
\end{equation}
Hence,with respect to the possibility of detecting squeezing, the filter bandwidth acts like a radiationless dephasing. Note also, that squeezing, which is lost through dephasing, cannot be recovered by optical filtering.

It should be noted at this point, that in all calculations we neglect the effect of back action of light reflected by the optical filter, compare~\cite{PhysRevA.42.503}. This means, we assumed the spectral filters to be slightly tilted with respect to the light to be measured, to suppress effects of the fields reflected from the filter to interfere with the original signal field. In turn, the reason for the reduction of squeezing is not due to back action in this scenario. Spectral filters, which are narrow compared to the squeezing spectrum of the signal field, act like delta functions reproducing the original field under convolution. Therefore, narrow optical filters, while diminishing the intensity of the light field substantially, are better for detecting squeezing.

\begin{figure}[h]
\includegraphics[width=7.5cm]{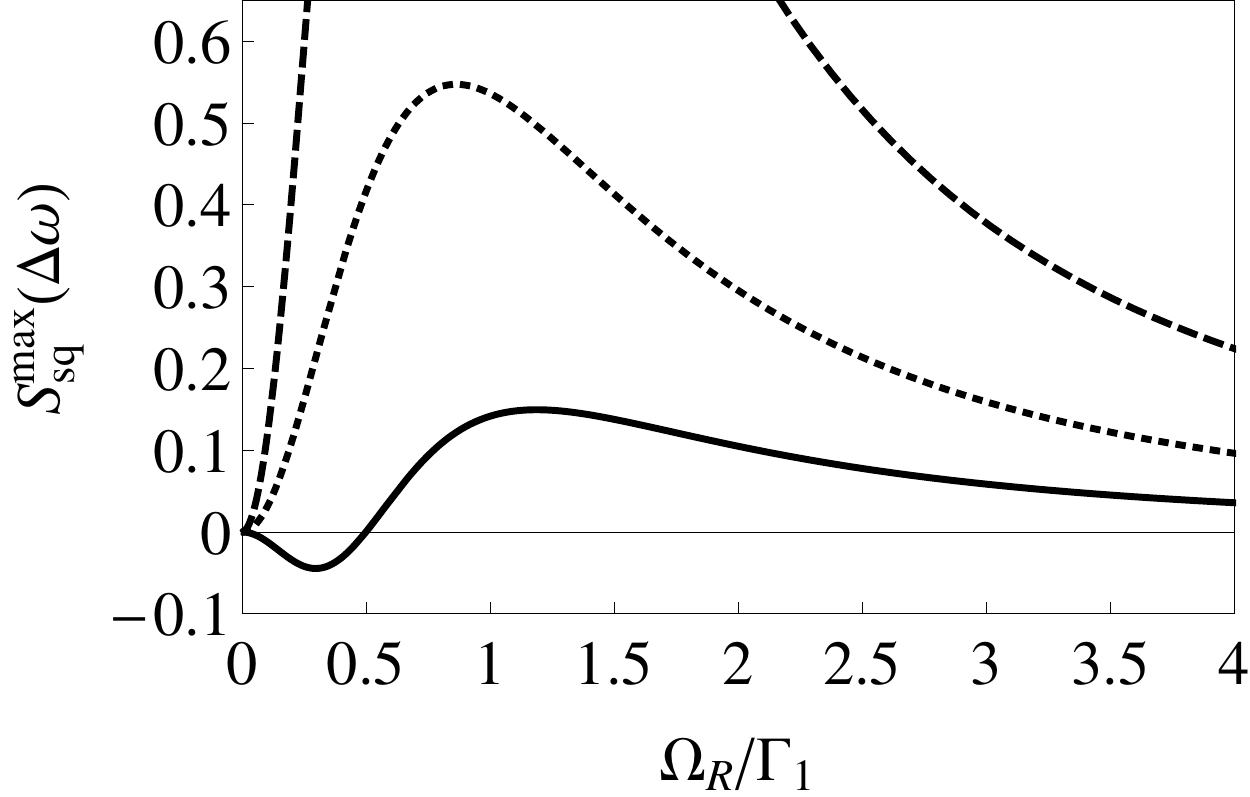}
\caption{\label{correlation_filter_2} The spectrum $S_{\text{sq}}^{\text{max}}$ for $\Delta\omega=0$ and different values of the pass-band width $\Gamma_\text{f}/\Gamma_1$: $1/3$ (dotted), $1/10$ (dashed), $0$ (solid).}
\end{figure}
In Fig.~\ref{correlation_filter_2} we show the squeezing spectrum at $\Delta\omega=0$. Applying Eq.~(\ref{eq.maxsq}) to the case of idealized filters, we obtain for the maximal squeezing at $\Delta\omega=0$
\begin{equation}
   S_{\text{sq}}^{\text{max}}(0)=\frac{2\sigma_{22}\Gamma_1}{\pi}\frac{\Gamma_1\Gamma_2+2\Omega_\text R^2-\Gamma_1^2}{[\Gamma_1\Gamma_2+\Omega_\text R^2]^2}.\label{eq.Sqmax0}
\end{equation}
For $\Gamma_1>\Gamma_2$, there are values of the driving $\Omega_\text R$, for which squeezing can be observed. However, if we include a nonzero filter width $\Gamma_\text f$, no squeezing occurs at all at $\Delta\omega=0$, as 
\begin{equation}
   S_{\text{sq}}^{\text{max}}(0)=\frac{2\sigma_{22}}{\pi\Gamma_\text f}\frac{(\Gamma_1+\Gamma_\text f)^2}{(\Gamma_1+\Gamma_\text f)(\Gamma_2+\Gamma_\text f)+\Omega_\text R^2}.\label{eq.Sqmax0f}
\end{equation}
As it is also seen from Fig.~\ref{correlation_filter_1}, for a nonzero filter bandwidth one also needs nonzero $\Delta \omega$ values to observe some nonclassical effect. 

\begin{figure}[h]
\includegraphics[width=7.5cm]{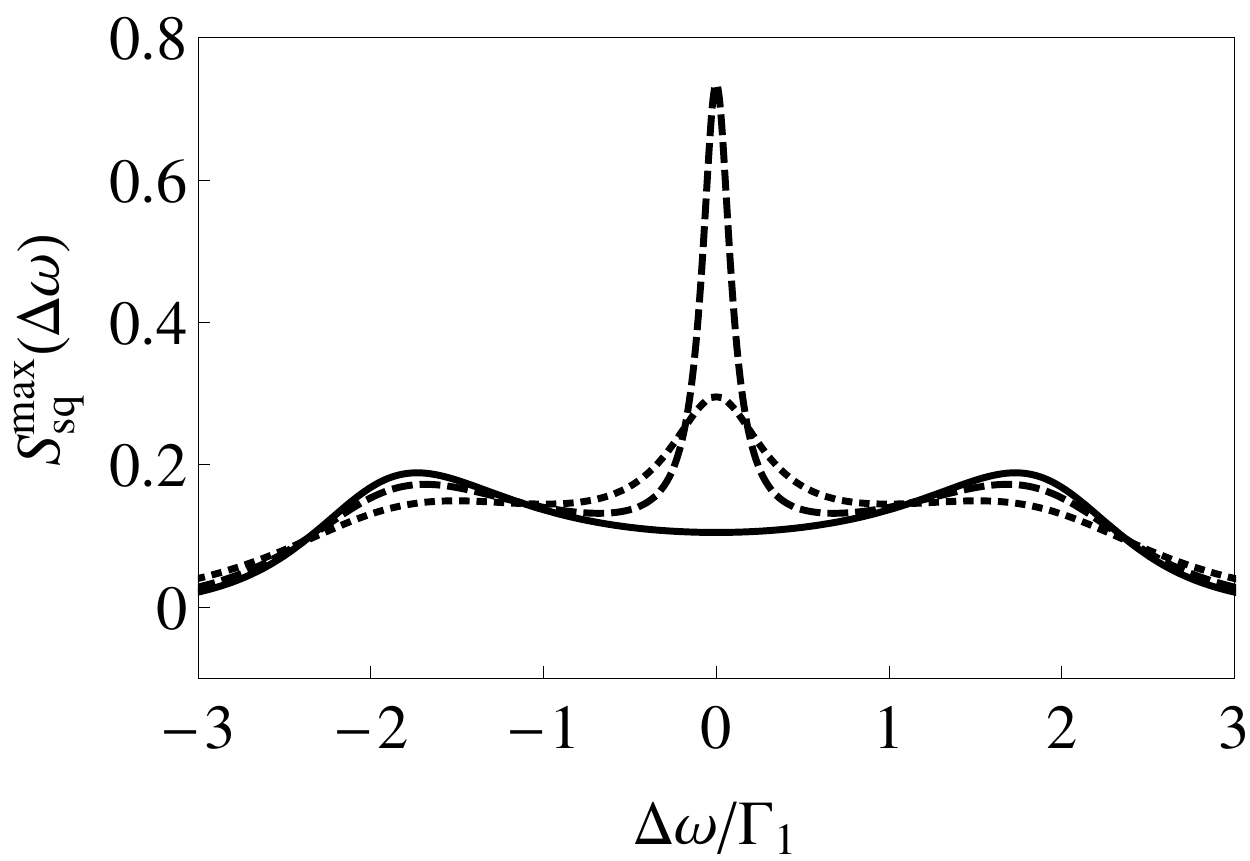}
\caption{The squeezing spectrum $S_{\text{sq}}^{\text{max}}$ for the high-driving-field limit ($\Omega_{\rm R}^2/\Gamma_1^2{=}4$) for different values of the optical filter pass-band width $\Gamma_\text{f}/\Gamma_1$: $1/3$ (dotted),  $1/10$ (dashed), $0$ (solid).}\label{fig.strong}
\end{figure}
The peak arising at $\Delta\omega=0$ for nonzero $\Gamma_\text f$ can also  be seen in the strong-driving-field limit $\Omega_{\rm R}^2{\gg}\frac{1}{2}\Gamma_1^2$ in Fig.~\ref{fig.strong}. Additionally, the two Mollow sideband peaks are visible in such a scenario. As discussed in Ref.~\cite{Zhou} a very similar effect was observed in the emission spectrum from a two-level atom driven by a strong coherent field to which an appropriate noise has been added.

\subsection{The squeezing spectrum for current filtering}
We now turn to the discussion of the current filtering procedure for squeezed light from resonance fluorescence. We use the scheme of Fig.~\ref{twin_doubleport} with Lorentz-type filters, for which the filter frequencies are chosen symmetrical relative to the laser frequency $\omega_{\rm L}$ or, equivalently, to the resonance frequency $\omega_0$ of the signal field. The squeezing spectrum is calculated as follows. One first calculates the total squeezing spectrum $\mathcal{S}$ by using Eq.~(\ref{two_filters_for_current}), for the filtering of the photocurrents as in the scheme of Fig.~\ref{twin_doubleport}. To compare with typical experimental procedures, in a next step the signal field is switched off and the corresponding correlations give the photon shot noise spectrum $\mathcal{S}_{\rm sn}$. The difference $\mathcal{S}_{\rm sq}{=}\mathcal{S}{-}\mathcal{S}_{\rm sn}$ is the squeezing spectrum of resonance fluorescence as it would be determined in an experiment. It is derived by specifying Eq.~(\ref{two_filters_for_current}) for Lorentzian filter functions with equal bandwidths $\Gamma_{\rm c}$ and setting frequencies $\omega_{{\rm c}_j}$ $(j=1,2)$, cf.~Appendix B. By varying the phases $\phi_1$ and $\phi_2$ of the local oscillator, one can reach the maximum squeezing effect, $\mathcal{S}_{\rm sq}{=}\mathcal{S}^{\rm max}_{\rm sq}$. Here we consider the case of resonance between the fluorescence and the mean  detection frequency.

\begin{figure}[h]
\includegraphics[width=7.5cm]{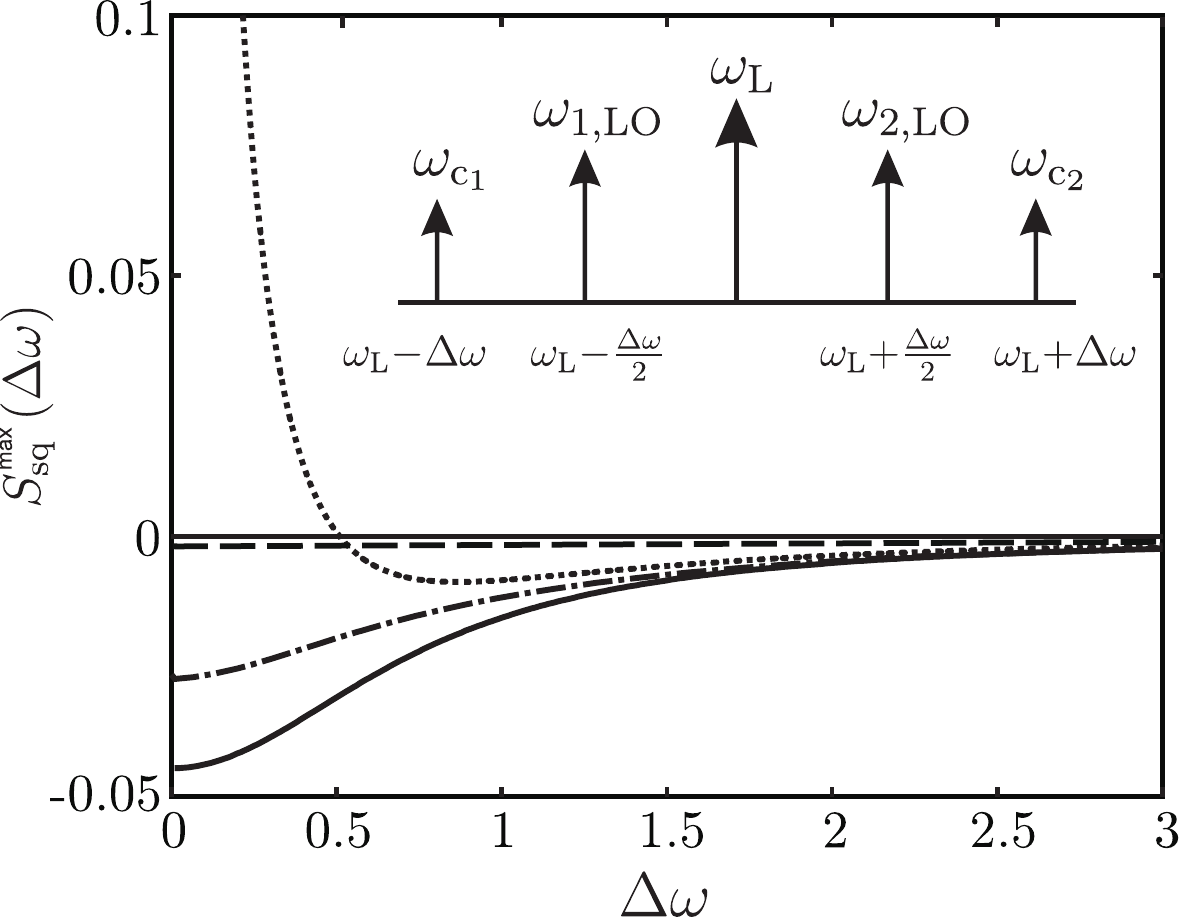}
\caption{The squeezing spectrum $S_{\text{sq}}^{\text{max}}$ obtained from filtered photocurrents (dashed-dotted: $\Gamma_{\text c}=1/10\Gamma_1$, dashed: $\Gamma_{\text c}=10\Gamma_1$), compared to optical filtering with $\Gamma_{\text f}=1/10\gamma$ (dotted) and ideal squeezing spectrum (solid). The spectra are calculated for $\Omega_{\text R}^2/\Gamma_1^2{=}1/12$. The inset shows the position of the central filter frequencies $\omega_{\text c_j}$ and local oscillator frequencies $\omega_{j,\text{LO}}$, $j=1,2$  with respect to the laser frequency $\omega_{\text L}$.}\label{fig.currentfilter}
\end{figure}

Unfortunately, the filtered squeezing spectrum cannot be given in a closed form for the current filtering as it was possible for the optical filtering in Eq.~(\ref{eq.maxsq}). From the analysis of Fig.~\ref{fig.currentfilter} it is evident that the current filtering with narrow-band filters is more suitable for the detection of squeezing than the radiation filtering with the same bandwidth parameters. This is especially evident for frequencies $\Delta\omega$ close to zero (compare the dashed-dotted with the dotted line). We conclude that the current filtering procedure is more suitable for analyzing the squeezing properties of light than the schemes involving the spectral filters.

For broadband current filters we obtain a flat squeezing spectrum, indicating a small observed squeezing effect only. Thus, we conclude that the narrow-band current filtering procedure is the most appropriate among other possibilities considered in the present paper. We also note that the relative positions of the local oscillator and current filter frequencies as indicated in the inset of Fig.~\ref{fig.currentfilter} are optimized for the detection of squeezing. For other possible frequencies one obtains squeezing effects in a very small $\Delta\omega$ range close to zero.

\section{Summary and Conclusions}\label{summary}
Based on the method of balanced homodyne correlation measurements we have studied the influences of the radiation field- and the photocurrent filtering on spectral correlation measurements of general quantum correlations of light. We have considered in detail the two different spectral measurement schemes for second-order field correlation functions and derived the connection between the original signal fields and the filtered field correlations, which are eventually detected. The theory has been formulated for normal- and time-ordered correlation functions of second order in the field operators. 

The general results have been illustrated for the example of the squeezing spectra of the resonance fluorescence of a two-level atom. Both the filtering of the radiation field and the filtering of the photocurrent have been analyzed. For the latter technique the optimal setting of the local oscillator and the current filter frequencies have been determined. 
Optical filtering substantially limits the available squeezing, that can be detected. Only for different filter resonance frequencies and very small filter bandwidth, squeezing can be observed. On the other hand, the current filtering is a powerful technique to analyze spectral correlation effects for the considered example of squeezing in atomic resonance fluorescence. In particular, it has been demonstrated that the current filtering scheme is better suited for the measurement of squeezing than the setup with optical spectral filters. This feature, together with the lower costs and better controllability of current filters in comparison with the spectral ones, makes current filtering more favorable for the experimental study of nonclassical light.

\section*{Acknowledgments}

PG, DV and WV gratefully acknowledge support by the Deutsche Forschungsgemeinschaft (DFG) through SFB 652. DV has benefited from discussions with D.~Karnaushenko, M.~de Oliveira and A.A.~Semenov. DV also acknowledges the  support  by the project N 0113U001093 of the National Academy of Sciences of Ukraine.

\appendix 
\section{Lorentzian radiation filter}\label{AppA}
Let us consider the measurement scheme in Fig.~\ref{twin_doubleport_filter} with two filters ${\rm SF}_j$, $j=1,2$. 
In order to simulate the action of the filter on the incident light field, we apply the special case of a Lorentzian filter function, which is a very typical elementary filter type. The shape of the function for the $j$th filter in the time domain reads as
\begin{equation}\label{filter_function}
T_{\text{f}_j}(t)=\Theta(t)\Gamma_\text{f} e^{-\Gamma_\text{f}t-i\omega_{\text{f}_j}t},
\end{equation}
where $\omega_{\text{f}_j}$ is the characteristic frequency of the $j$th filter and $\Gamma_\text{f}$ the
pass bandwidth, which is the same for both filters. The unit step function  $\Theta(t)$ ensures causality. One obtains after the substitution in  Eq.~(\ref{eq.fspectral11}) the following expression for the function $F^{(1,1)}_\text{spectral}$

\begin{equation}
\begin{split}
	&F^{(1,1)}_\text{spectral} = E_{\rm LO}^2\Gamma_{\rm f}^2\int_0^t\!\! d  t_1'\int_0^t\!\! d  t_2'e^{-\Gamma_{\text f}(2t-t_1'-t_2')}\\
	&\quad\times\left\langle\tno\prod_{j=1}^2\Bigg[\hat{\mathcal E}^{(+)}(t_j')e^{-i\omega_{\text f_j}t_j'}e^{i(\omega_{j,{\rm LO}}-\omega_{\text f_j})t-i\varphi_j}\right.\\
	&\qquad+\left.\hat{\mathcal E}^{(-)}(t_j')e^{i\omega_{\text f_j}t_j'}e^{-i(\omega_{j,{\rm LO}}-\omega_{\text f_j})t+i\varphi_j}\Bigg]\tno\right\rangle.
\end{split}
\end{equation}

For the case of both filters having the same central pass frequency as the respective phase shifted $\rm LO$-fields, $\omega_{j,\text{LO}}=\omega_{\text{f}_j}$, we obtain
\begin{equation}
\begin{split}
&\left\langle\!\tno\! \hat{\tilde E}_1^{(+)} \hat{\tilde E}^{(+)}_2
\! \tno\!\right\rangle =\Gamma_\text{f}^2\! \int_0^t d  t_1'\!\! \int_0^t d  t_2'
e^{-\Gamma_\text{f}(2t{-}t_1'{-}t_2')}\\
&\qquad\qquad \times e^{i(\omega_{\text{f}_1}t_1' {+} \omega_{\text{f}_2}t_2')}\bigg\langle\!\tno\!\hat{\mathcal{E}}^{(+)}(t_1')\hat{\mathcal{E}}^{(+)}(t_2')
\!\tno\!\bigg\rangle.
\end{split}
\end{equation}
Further factorizing the incoming fields into slowly varying amplitude and fast oscillating term with mean frequency $\omega_0$,
\begin{equation}\label{slow}
\hat{\mathcal E}^{(\pm)}=\hat{\tilde{\mathcal E}}^{(\pm)}e^{\mp i \omega_0t},
\end{equation}
and denoting the frequency difference by $\Delta \omega {=} \omega_{\text{f}_2}{-}\omega_{\text{f}_1}$  we get in terms of
new variables $\tau_j {=} t{-}t_j'$ for stationary fields the following expression
\begin{equation}
\begin{split}
&\left\langle\!\tno\! \hat{\tilde E}_1^{(+)} \hat{\tilde E}^{(+)}_2
 \! \tno\!\right\rangle =\Gamma_\text{f}^2\!\int_0^t d  \tau_1 \!\!\int_0^t d \tau_2
 e^{-\Gamma_\text{f}(\tau_1{+}\tau_2)}\\
&\qquad\times e^{-i\frac{\Delta\omega}{2}(\tau_2 - \tau_1)} \left\langle\!\tno\! \hat{\tilde{\mathcal E}}^{(+)}(\tau_2{-}\tau_1)\hat{\tilde{\mathcal E}}^{(+)}(0)
  \!\tno\!\right\rangle.
\end{split}\label{E+E+F}
\end{equation}
Denoting  $\tau{=}\tau_2{-}\tau_1$ and integrating this expression over $\tau'{=}\tau_2{+}\tau_1$ 
 yields
\begin{equation}\label{plusplus-correlation}
\begin{split}
\left\langle\!\tno\! \hat{\tilde E}_1^{(+)} \hat{\tilde E}^{(+)}_2
\! \tno\!\right\rangle&{=}\frac{\Gamma_\text{f}}{4} (1-e^{-\Gamma_{\rm f}t})\int_{0}^{t} d  \tau e^{{-} i\Delta\omega\tau/2}\\
&\qquad\times\left\langle\!\tno\! \hat{\tilde{\mathcal E}}^{(+)}(\tau)\hat{\tilde{\mathcal E}}^{(+)}(0)
\!\tno\!\right\rangle.
\end{split}
\end{equation}
The negative-negative frequency correlation function is obtained from (\ref{plusplus-correlation}) by conjugation.  In the case of negative-positive correlation we obtain 
\begin{equation}\label{spectral-minusplus-correlation}
\begin{split}
&\left\langle\!\tno\! \hat{\tilde E}_1^{(-)} \hat{\tilde E}^{(+)}_2\!\tno\!\right\rangle = \Gamma_\text{f}^2\!\int_0^t\!\! d  t_1'\!\! \int_0^t\!\! d  t_2'
e^{-\Gamma_\text{f}(2t{-}t_1'{-}t_2')}\\
&\qquad\times e^{i\Delta\omega(t_1' {+} t_2')}
\Big\langle\tno \hat{\tilde{\mathcal E}}^{(-)}(t_1')\hat{\tilde{\mathcal E}}^{(+)}(t_2')\tno\Big\rangle.
\end{split}
\end{equation}
Using Eqs~(\ref{plusplus-correlation}), (\ref{spectral-minusplus-correlation}) one can calculate the squeezing spectrum of the filtered light.

\section{Lorentzian current filter} \label{AppB}
We consider the two filter frequency setup for current filtering from Fig.~\ref{twin_doubleport}.
For the current filter and for the detector response functions the same Lorentz-type functions as for the spectral filter [cf.~(\ref{filter_function})] are used. Namely, we assume, that
\begin{eqnarray}\label{eq.TS}
T_{\text c_j}(t)&=&\Theta(t)\Gamma_{\text{c}}e^{-\Gamma_{\text{c}} t {-} i\omega_{\text{c}_j}t},\\
S(t)&=&\Theta(t)\Gamma_{\text{s}}e^{-\Gamma_{\text{s}} t {-} i\omega_\text{s}t},
\end{eqnarray}
where $\omega_{\text c_j}$ ($j=1,2$), and $\omega_{\text s}$ are correspondingly current filter and detector response frequencies and $\Gamma_{\text c}$, $\Gamma_{\text s}$ are the pass-band widths of the current filter and detector, respectively. Substituting Eq.~(\ref{eq.TS}) into Eq.~(\ref{two_filters_for_current}) we arrive at
\begin{equation}\label{eq.F12}
\begin{split}
&F^{(1,1)}_\text{current}=\Gamma_{\text c}^2\Gamma_{\text s}^2 N^2 E_{\rm LO}^2\int_0^t d  t_1'\int_0^t d  t_2'\\
&\quad \times e^{-\Gamma_{\text c}(2t-t_1'-t_2')}e^{-i\omega_{\text c_1}(t-t_1'){-}i\omega_{\text c_2}(t-t_2')}\\
&\quad\times\bigg\langle\tno \prod_{j=1}^2\int^{t_j'+\Delta t}_{t_j} \hspace{-0.5cm}  d \tau_j\!
 \int^{t_j'+\Delta t}_{t_j'} \hspace{-0.5cm}  d \tau'_j 
 e^{-\Gamma_{\text{s}}( \tau_j{-}\tau'_j)}\\
&\qquad\times \bigg[ \mathcal{\hat{\tilde E}}^{(-)}(\tau_1)e^{-i(\omega_\text{s}-\omega_0)\tau_j }e^{i(\omega_\text{s}-\omega_{j,\text{LO}})\tau_j' {+}i \varphi}\\
&\qquad+\mathcal{\hat{\tilde E}}^{(+)}(\tau'_2)e^{i(\omega_\text{s}-\omega_0)\tau_j' }e^{-i(\omega_\text{s}-\omega_{j,\text{LO}})\tau_j {-}i \varphi}\bigg] \tno\bigg\rangle,
\end{split}
\end{equation}
where we have used the  slowly varying amplitudes,
$\hat{\mathcal E}^{(\pm)}{=}\hat{\tilde{\mathcal E}}^{(\pm)}e^{\mp i \omega_0 t}$.
For the resonance condition $\omega_0=\omega_\text{s}$ Eq.~(\ref{eq.F12}) can be further simplified.

%\bibitem{PRA.36.3803} L. Kn\"oll, W. Vogel \& D.-G. Welsch, Phys. Rev. A, \textbf{36}, 3803 (1987)


\begin{thebibliography}{18}
\bibitem{Edmund} \textsl{One may look into any current catalogue of scientific optical components to study the various optical filters. For example: Edmund Optics, Newport, etc.}
\end{thebibliography}
\end{document}